\documentclass[twocolumn]{aastex701}
\usepackage{CJK}

\newcommand{\Fe}{{\rm Fermi\mbox{-}LAT}}
\newcommand{\TeV}{{\rm TeV}}
\newcommand{\QPO}{{\rm QPO}}
\newcommand{\PSD}{{\rm PSD}}

\shorttitle{Credibility of Gamma-Ray QPO Candidates}
\submitjournal{ApJ}

\begin{document}
\begin{CJK*}{UTF8}{gbsn}

\title{Assessing the Credibility of Gamma-Ray \QPO\ Candidates in 41 \TeV-Selected Blazars}

\shortauthors{Yang et al.}

\author[0000-0003-3474-1125]{Wen-Xin Yang ({\Unicode{103}{104}\Unicode{101}{135}\Unicode{153}{168}})}
\affiliation{Center for Astrophysics, Guangzhou University, Guangzhou, Guangdong, 510006, China.}
\affiliation{Dipartimento di Fisica e Astronomia ``G. Galilei'', Universit\`a di Padova, Padova, 35131, Italy.}
\affiliation{Astronomy Science and Technology Research Laboratory of Department of Education of Guangdong Province, Guangzhou 510006, China.}
\affiliation{Greater Bay Brand Center of the National Astronomical Data Center, Guangzhou 510006, China.}
\email{2112019028@e.gzhu.edu.cn}

\author[0000-0003-3863-9777]{Yi Liu ({\Unicode{82}{24}\Unicode{96}{33}})}
\affiliation{Center for Astrophysics, Guangzhou University, Guangzhou, Guangdong, 510006, China.}
\affiliation{Astronomy Science and Technology Research Laboratory of Department of Education of Guangdong Province, Guangzhou 510006, China.}
\affiliation{Greater Bay Brand Center of the National Astronomical Data Center, Guangzhou 510006, China.}
\email[show]{pinux@gzhu.edu.cn}

\author{Hong-Guang Wang ({\Unicode{115}{139}\Unicode{109}{42}\Unicode{81}{73}})}
\affiliation{Center for Astrophysics, Guangzhou University, Guangzhou, Guangdong, 510006, China.}
\affiliation{Astronomy Science and Technology Research Laboratory of Department of Education of Guangdong Province, Guangzhou 510006, China.}
\affiliation{Greater Bay Brand Center of the National Astronomical Data Center, Guangzhou 510006, China.}
\email{hgwang@gzhu.edu.cn}

\author{Denis Bastieri}
\affiliation{Dipartimento di Fisica e Astronomia ``G. Galilei'', Universit\`a di Padova, Padova, 35131, Italy.}
\affiliation{Center for Astrophysics, Guangzhou University, Guangzhou, Guangdong, 510006, China.}
\email{denis.bastieri@unipd.it}

\author[0000-0002-5929-0968]{Jun-Hui Fan ({\Unicode{106}{10}\Unicode{81}{155}\Unicode{143}{137}})}
\affiliation{Center for Astrophysics, Guangzhou University, Guangzhou, Guangdong, 510006, China.}
\affiliation{Astronomy Science and Technology Research Laboratory of Department of Education of Guangdong Province, Guangzhou 510006, China.}
\affiliation{Greater Bay Brand Center of the National Astronomical Data Center, Guangzhou 510006, China.}
\email[show]{fjh@gzhu.edu.cn}

\begin{abstract}
We analyze quasi-periodic oscillation (\QPO) candidates in the 0.1--100 GeV \Fe\ light curves of 41 \TeV\ blazars with a conservative, multi-stage timing analysis combining FFT power spectra, red-noise Monte Carlo simulations, false-discovery-rate control, bootstrap period uncertainties, split-sample cross-validation, Lomb--Scargle and WWZ diagnostics, a detection-only sensitivity test, observing-window checks, and injection--recovery experiments. A first-pass FFT search flags 16 sources above a raw 95\% level and six strong candidates after data-quality and split-sample filtering, none satisfying a fully automated multi-method tier. Source-level significance is quantified with forward simulations that preserve the observed sampling and non-Gaussian flux distribution, treating the red-noise slope as a nuisance parameter and never assuming independence of periodogram powers. At simulation-selected reference slopes, only two of the 41 sources reach a raw source-level $p<0.05$, only J1555.7+1111 reaches Benjamini--Hochberg significance, and none survives Benjamini--Yekutieli correction. No source survives the conservative slope envelope. Even J1555.7+1111, which the first-pass search flagged as significant but did not rank among its strong candidates, has a global $p$-value that rises to $7.5\times10^{-3}$ under the least favorable slope ($\sim0.31$ after the full-sample search), so it is not a robust detection. With substantial period uncertainties, a weak observing window, and limited injection--recovery sensitivity, we establish no robust gamma-ray \QPO\ detection: a sampling-faithful red-noise treatment substantially contracts the candidate set, providing an explicit assessment of \QPO\ credibility.
\end{abstract}


\section{Introduction}
\label{sec:intro}

Blazars are among the most strongly variable active galactic nuclei and exhibit broadband variability over a wide range of timescales \citep{urry1995blazar,padovani2017blazar}. For that reason, claims of quasi-periodic oscillations (\QPO\ signals) in blazar light curves have attracted sustained attention. If real, such timescales could constrain processes such as jet precession, geometric modulation, recurrent structures in relativistic outflows, or variability linked to the accretion system \citep{camenzind1992helical,villata1999helical,spada2001shock,liska2018jet,frank2002accretion}. A number of source-specific gamma-ray studies have therefore reported candidate periodic behavior in individual blazars \citep[e.g.,][]{tavani2018pg1553,sandrinelli2014qpo2155,ackermann2015qpo,zhou2018pks2247}, and broader survey-style searches have also begun to revisit the same question in a homogeneous way \citep{penil2020periodicity,ren2023qpoagn}.

At the same time, gamma-ray \QPO\ claims are statistically difficult to establish. Blazar light curves are red-noise dominated, often contain long stretches of low-significance bins, and are observed over finite baselines with coarse monthly sampling. Under these conditions, apparent peaks can arise from stochastic variability, limited frequency resolution, or differences in how individual period-search methods localize broad structures in the power spectrum \citep{abdo2010psd,vaughan2005rednoise,vaughan2016rednoise}. The central difficulty is therefore not simply to find a peak, but to determine whether a reported timescale remains credible once the null model, sampling limitations, and method dependence are treated explicitly.

This motivates the main scientific question of the present work: when a uniform sample of \TeV-selected blazars is analyzed with a conservative and reproducible timing workflow, how many gamma-ray \QPO\ candidates remain plausible after successive statistical and robustness tests? We address this question with monthly 0.1--100 GeV \Fe\ light curves for 41 \TeV\ blazars. Rather than treating an isolated periodogram maximum as a detection, we evaluate each candidate in stages: source-specific red-noise significance, multiple-testing correction, data-quality filtering, a split-sample half-to-half consistency screen, independent-method comparison, detection-only reanalysis, and source-level follow-up diagnostics.

The paper is organized as follows. Section~\ref{sec:data} describes the archived \Fe\ light-curve sample and its basic properties. Section~\ref{sec:methods} presents the timing analysis, significance tests, candidate hierarchy, and robustness diagnostics. Section~\ref{sec:results} reports the survey results, the contraction of the candidate set under progressively stricter criteria, and the source-level follow-up outcomes. Section~\ref{sec:discussion} discusses the implications of the null and ambiguous cases, the role of sensitivity limitations, and the physical interpretation of the remaining candidate timescales. Section~\ref{sec:conclusions} summarizes the main conclusions.

 \section{Sample and Data}
\label{sec:data}

The sample comprises 41 \TeV-selected blazars with monthly \Fe\ light curves in the 0.1--100 GeV band. These sources are selected for being bright at \TeV\ energies and well detected in the GeV band \citep{abdollahi2020fermi4fgl}; the light-curve data are taken from \citet{wang2024}. The catalog provides 178 uniformly sampled 30-day bins per source over Modified Julian Date (MJD) 54683--59993, corresponding to a 5310 day baseline. A full source list with timing and detection-quality quantities is provided in Appendix~\ref{sec:appendix_sample}, and data-access details are given in the Data and Code Availability statement.

\subsection{Light-curve Construction}

Each source is represented by a photon flux and an associated uncertainty for every time bin. The monthly light curves are products of the standard \Fe\ analysis chain, with source fluxes computed using standard event-selection and quality-control procedures as described in \citet{wang2024}.

The use of a uniform 30-day cadence is a practical compromise between temporal resolution and photon statistics. Shorter bins would increase the number of nominal cycles available for period searches, but they would also produce a larger fraction of low-significance measurements for many sources. Longer bins would reduce noise at the cost of smearing candidate variability on sub-year timescales. The adopted cadence is therefore intended for a survey-level search for broad recurrence timescales rather than for resolving fine phase structure.

\subsection{Data-quality Characterization}

Because low-flux gamma-ray bins may behave more like upper limits than secure detections, we mark bins with ${\rm Flux}<3\times{\rm Error}$ as low-significance measurements. The mean detection fraction across the sample is 61.5\%, and 18 sources have detection fractions below 50\%. This fact becomes important when interpreting candidate periods: a narrow power-spectrum peak from a source with poor detection fraction should not be treated on the same footing as a peak from a well-detected source.

In the present workflow these low-significance bins are retained in the uniformly sampled light curves, but their impact is tracked explicitly through the detection-fraction statistic used later in candidate grading. This choice preserves the regular time grid needed for the Fourier analysis while recognizing that not all sources contribute equally informative data across the full 14.5-year baseline.

\subsection{Sample Composition}

The sample identifies a mixed \TeV\ blazar population containing six flat-spectrum radio quasars (FSRQs), eleven high-frequency-peaked BL Lac objects (HBLs), fourteen intermediate-frequency-peaked BL Lac objects (IBLs), and ten blazar candidates of uncertain subclass (BCUs), as summarized in Table~\ref{tab:appendix_sample}. No source carries a secure low-frequency-peaked BL Lac (LBL) label in the published classification. Broad source-type comparisons are therefore treated as exploratory only.

Redshift information is not uniformly available for all sources in this sample, so we do not use redshift-dependent quantities in the candidate ranking itself. The sample table in the Appendix is correspondingly limited to source identity, subtype labels where available, baseline coverage, and detection-quality statistics.

 \section{Methods}
\label{sec:methods}

\subsection{Timing Analysis}

For each source, we subtract the mean flux, linearly interpolate missing values if needed, apply a Hann (Hanning) window, and compute the discrete Fourier or FFT power spectrum. The taper is used in the standard sense of spectral analysis to suppress leakage from strong low-frequency structure into neighboring frequencies \citep{harris1978windows}. Candidate peaks are searched only over periods between 60 and 1000 days, corresponding to the reliable range for the monthly cadence and available baseline. Subtracting the sample mean introduces an additional uncertainty of order $\sigma_{\rm mean}\approx \sigma_{\rm flux}/\sqrt{N_{\rm eff}}$, where $N_{\rm eff}$ is the effective number of detected bins contributing to the source mean. For the present monthly light curves, $N_{\rm eff}$ is typically of order $10^2$, so this extra term is only at the $\sim$8\%--10\% level of the characteristic single-bin scatter and is therefore negligible compared with the bin-level flux uncertainties already carried by the light curves.

The lower bound of 60 days is set by the effective Nyquist scale for 30-day sampling, while the upper bound of 1000 days is chosen to avoid interpreting features too close to the full data baseline as well-resolved periodic signals. Within this search window, the dominant candidate period for each source is defined by the maximum power. As a direct preprocessing check, we also compared the six strong candidates against Lomb--Scargle periodograms computed from the original, non-mean-subtracted time series on the same Fourier frequency grid. Across this strong-candidate subset, the dominant peaks and their relative prominence agree at the level needed for source ranking, which indicates that the interpolation and zero-mean steps do not drive the main candidate selection. At this stage the Fourier spectrum is therefore used primarily as a ranking device, and the downstream significance and stability tests determine whether the highest peak remains credible.

\subsection{Red-noise Significance Testing}

For each source we model the red-noise continuum as a single power law, $P(f)\propto f^{-\alpha}$, and we assess significance entirely through forward simulation. We deliberately do not estimate the slope by an ordinary least-squares regression of $\log_{10} P(f)$ on $\log_{10} f$, where $P(f)$ is the periodogram power at temporal frequency $f$: periodogram ordinates are approximately $\chi^2$-distributed rather than Gaussian, and for the irregularly sampled, detection-only light curves used in the significance test they are not mutually independent, so a log--log least-squares fit is not a statistically valid estimator of the slope \citep{vaughan2005rednoise,vio2010unevenly}.

Instead, we calibrate significance with simulated light curves that reproduce both the assumed power-law \PSD\ and the observed, generally non-Gaussian, flux distribution, following the algorithm of \citet{emmanoulopoulos2013lightcurves}, which extends the Gaussian scheme of \citet{timmer1995rednoise} to match the empirical flux probability density function. Each realization is propagated through the identical sampling window, frequency grid, and peak statistic as the data: we apply the observed detection mask (${\rm Flux}>3\times{\rm Error}$), evaluate the same period-search statistic over the same 60--1000 day grid, and record the maximum simulated peak power. The source-level global $p$-value is the fraction of the $N_{\rm sim}=10^{4}$ realizations whose maximum peak power equals or exceeds the observed maximum, evaluated as $(N_{\rm exceed}+1)/(N_{\rm sim}+1)$ to avoid artificial zero probabilities, where $N_{\rm exceed}$ is the number of realizations exceeding the observed peak. Because the search over trial periods within a source is folded into this maximum-peak comparison, the resulting $p$-value already incorporates the look-elsewhere effect within that source. As a sensitivity check on the fixed observed mask, we also repeat the calculation for the three lowest-$p$ sources while recomputing ${\rm Flux}>3\times{\rm Error}$ separately in every simulated realization.

Deriving significance from complete simulated light curves that share the observed sampling is central to our response to the statistical concerns that arise for irregularly sampled data. Since each null realization passes through the same window function as the data, the full joint distribution of periodogram powers---including all covariance introduced by the sampling---is reproduced empirically in the simulated ensemble. The test therefore never assumes that neighbouring powers are statistically independent and never multiplies single-frequency false-alarm probabilities, which directly addresses the correlation of periodogram ordinates under uneven sampling \citep{vio2010unevenly}. This simulation-based calibration is the approach advocated for red-noise periodicity searches by \citet{vaughan2010bayesian} and \citet{emmanoulopoulos2013lightcurves}.

This procedure yields one p-value for each source because the observed highest peak is compared with the distribution of highest peaks from the simulated null ensemble. In other words, the search over many trial periods within that source is already built into the Monte Carlo comparison. The subsequent multiple-testing correction is therefore applied across sources rather than across trial periods within a source. This distinction matters because the analysis aims to control both false positives for individual sources and the overall credibility of claims made for the full 41-source sample. For the representative source panels shown later, we also display frequency-dependent 95\% and 99\% threshold curves derived from the same source-specific Monte Carlo ensembles, so that the observed \PSD\ and Lomb--Scargle power can be compared directly with the null expectation at each sampled period.

The red-noise slope $\alpha$ is only weakly constrained by short, irregularly sampled light curves, and we do not treat it as precisely known. For descriptive reference values, we select $\alpha$ by simulation-based minimum distance: at each grid value we compare the normalized observed period-search spectrum with the median spectrum from 100 forward simulations and choose the value that minimizes the mean squared log-spectrum distance. This is not an OLS fit or a likelihood-based confidence statement; it supplies the reference slopes used for the raw, BH, and BY counts reported below. For the inferential conclusion, we instead treat $\alpha$ as a nuisance parameter: for each source we recompute the global $p$-value on a grid of slopes spanning $0\le\alpha\le4$ in steps of $0.25$ and report a conservative envelope, defined as the largest (least significant) global $p$-value over this range. A candidate is regarded as robust only if it remains significant under this worst-case slope, so that our conclusions do not hinge on the simulation-selected reference value. The single-power-law null is intentionally simple: we do not fit a separate high-frequency white-noise plateau, because the monthly cadence and finite baseline leave too few high-frequency bins to constrain a two-component $Af^{-\alpha}+C$ model, with normalization $A$ and frequency-independent white-noise level $C$, across a uniform 41-source survey. This baseline is not intended as a full \PSD-shape inference engine, but the conservative treatment of $\alpha$ ensures that the candidate assessment is not dominated by the choice of slope.

\subsection{Multiple Testing and Candidate Tiers}

We report raw significance values, but all source-level claims are based on multiple-testing correction across the 41-source sample. We compute Bonferroni, Benjamini--Hochberg (BH), and Benjamini--Yekutieli adjusted values \citep{benjamini1995fdr,benjamini2001by} to account for the fact that many sources are tested in the same survey. Bonferroni is the most conservative option because it aims to keep the chance of making even one false positive claim across the full sample very small. BH is less restrictive because it instead controls the expected fraction of false positives among the sources declared significant. Benjamini--Yekutieli follows the same general idea as BH but remains valid under more general dependence among the tests, at the cost of greater conservatism. Our working significance threshold is BH $q<0.05$, where $q$ denotes the BH-adjusted p-value.

To avoid relying on significance alone, we define three automated evidence tiers and a final source-level follow-up layer:
\begin{enumerate}
\item \textbf{Statistically significant}: raw significance $\ge 95\%$ and BH $q<0.05$.
\item \textbf{Strong candidate}: statistically significant, detection fraction $\ge 50\%$, and split-sample half-to-half period classification of Consistent or Partial.
\item \textbf{Automated multi-method candidate}: a strong candidate that also satisfies quantitative support criteria from the auxiliary methods. Lomb--Scargle support requires BH-corrected Lomb--Scargle significance together with FFT-to-Lomb--Scargle period consistency at the Consistent or Partial level. WWZ support requires BH-corrected WWZ significance, a WWZ stability class of Persistent or Partial, and FFT-to-WWZ median-period consistency at the Consistent or Partial level.
\item \textbf{Source-level follow-up case}: a strong candidate that also survives a detection-only Lomb--Scargle reanalysis and is then examined individually through method-to-method period comparison, observing-window context, and injection--recovery diagnostics.
\end{enumerate}

The threshold of 50\% for detection fraction is pragmatic rather than absolute. It separates sources for which most of the baseline contributes actual detections from those for which a large fraction of the light curve is close to the detection threshold. The source-level follow-up layer is not a purely automatic category; instead, it reflects which sources remain sufficiently persuasive to justify detailed source-by-source discussion once low-significance bins are handled more conservatively and the period estimates are compared across methods. We therefore use these tiers as an interpretive scaffold rather than as immutable physical classes.

Because the 50\% detection-fraction cut is somewhat pragmatic, we also evaluate nearby thresholds of 40\% and 60\% as a sensitivity test on the strong-candidate count. This shows whether the strong-candidate tier is dominated by a single threshold choice.

\subsection{Uncertainty and Stability Checks}

We estimate period uncertainty with a moving-block bootstrap (500 resamplings; block size 5 bins) and characterize the resulting period spread through the standard deviation and 95\% interval. Block resampling is used because it preserves short-range temporal dependence more realistically than a pointwise bootstrap for correlated light curves \citep{kunsch1989bootstrap}. We also perform split-sample cross-validation by analyzing the first and second halves of each light curve independently. The resulting class is defined only from the ratio between the dominant periods recovered from the two half-light curves: if that ratio is below 1.2, we classify the source as Consistent; ratios from 1.2 to 2.0 are classified as Partial; larger disagreements are classified as Inconsistent/Undetermined. These thresholds are intended to distinguish modest shifts expected from finite resolution and broad peaks from larger disagreements more likely to indicate unstable or poorly localized timescales.

The block length of five bins corresponds to approximately 150 days and is intended to preserve short-range temporal correlations while still allowing a diverse set of resampled realizations. The bootstrap is not used as a formal detector; instead, it quantifies how sharply the candidate period is localized under resampling. The split-sample test serves a complementary role by asking whether a similar recurrence scale is recovered independently in the first and second halves of the baseline. It should be interpreted as a half-to-half agreement screen rather than as a requirement that either half reproduce the full-sample candidate period exactly. A source can therefore be highly significant and still fail the robustness filters if the signal is poorly localized in period or shifts strongly with time, while a source that passes this screen may still have half-sample peaks displaced from the full-sample optimum. For the split-sample check, an implicit source-level stability fraction is therefore the fraction of the two half-light-curve tests that remain within the adopted 20\% period-consistency tolerance; in practice we summarize this information more compactly through the Consistent/Partial/Inconsistent/Undetermined classes reported in the tables and figures.

Low-significance bins are examined through a separate detection-only sensitivity analysis. In that branch, bins with ${\rm Flux}<3\times{\rm Error}$ are removed rather than interpolated, where ${\rm Error}$ denotes the quoted flux uncertainty for that bin, and the surviving irregularly sampled time series is reanalyzed with Lomb--Scargle periodograms \citep{scargle1982} and source-specific red-noise Monte Carlo simulations. Lomb--Scargle is used here because it provides a standard period-search statistic for unevenly sampled data and therefore avoids reconstructing a censored regular grid. Its source-level significance is estimated from the distribution of maximum Lomb--Scargle powers recovered from the same source-specific Monte Carlo ensemble, evaluated on the same frequency grid as the data. Because neighboring Lomb--Scargle powers are not strictly independent, we treat these Monte Carlo thresholds as operational and slightly conservative rather than as an exact analytic false-alarm law. We intentionally do not interpret this test as a detection-only FFT or WWZ reanalysis, because those methods would require reconstructing a regular time series after censoring. The goal of this branch is therefore narrower: it asks whether the candidate timescale remains visible when only secure detections are retained.

\subsection{Ancillary Analyses}

In addition to the timing analysis, we derive descriptive time-domain and frequency-domain features and use them for a simple three-cluster characterization of variability behavior. The clustering is performed with k-means in standardized feature space \citep{macqueen1967kmeans}, and cluster compactness is summarized with silhouette widths \citep{rousseeuw1987silhouettes}. These features include detection fraction, summary moments of the flux distribution, signal-to-noise proxies, and power-spectrum descriptors. The clustering results are used only as descriptive context and not as evidence for or against \QPO\ reality. Similarly, the source-type comparisons presented later are exploratory and are retained only to show where the candidate set sits within the broader sample.

We also compute the spectral window from the actual observation times. The spectral window characterizes periodic structure introduced by the cadence itself, independent of the source flux values, and is therefore used as a check on whether a candidate timescale could be favored by sampling alone. In the present monthly sampled data set, the window structure inside the 60--1000 day search range is weak, so we use it only as a diagnostic check on the cadence rather than as a decisive source-ranking criterion. WWZ diagnostics are evaluated in the spirit of time-frequency localization methods for unevenly sampled series \citep{foster1996wwz}. In this context, WWZ complements FFT and Lomb--Scargle by asking not only where power is concentrated in period, but also whether that power persists coherently across the observing baseline. In the adopted configuration, the WWZ map spans periods of 60--1000 days on a grid of 60 trial periods and 20 time centers, with decay constant $c=0.0125$. This grid is intentionally moderate in resolution: it is fine enough to track broad ridges over the available baseline, but coarse enough to avoid over-interpreting any substructure that is not well constrained by monthly sampling. We summarize each map through the global peak power, the median period traced by the strongest ridge, and a stability fraction. The WWZ significance is estimated in the same simulation-based spirit as the FFT and Lomb--Scargle tests: for each source we recompute the WWZ peak statistic for red-noise Monte Carlo realizations drawn from the fitted source-specific null model and compare the observed peak against that ensemble. Here the WWZ stability fraction $f_{\rm stab}$ is defined as the fraction of sampled time centers for which the dominant ridge period remains within 25\% of the median ridge period for that source. Stability fractions $\ge 0.7$ are classified as Persistent, those in the range $0.4\le f_{\rm stab}<0.7$ as Partial, and lower values as Weak.

Finally, injection--recovery experiments are performed by adding sinusoidal signals to red-noise realizations and then recomputing the detection pattern from the simulated fluxes and uncertainties. By ``red-noise realizations'' we mean fully synthetic light curves generated with the \citet{emmanoulopoulos2013lightcurves} procedure, which reproduces both the power-law \PSD\ and the observed flux distribution, evaluated at the conservatively adopted source-specific slope; we do not inject the sinusoid directly into the observed light curves. This allows the effective detection fraction to respond to the injected signal instead of being fixed to the observed mask. For the source-level follow-up cases, the injected period is tied to the timescale that survives into the source-level discussion rather than being fixed a priori to the FFT peak. The injected amplitude is reported in units of the observed light-curve standard deviation $\sigma$, measured from the source-centered monthly flux series. A recovery is counted as successful only when the recovered period falls within the adopted tolerance and the recovered peak also exceeds a source-specific 95\% null threshold for that method. Throughout the paper, ``period agreement within the adopted tolerance'' means that the ratio between the injected and recovered periods is $\le 1.2$, corresponding to agreement within 20\% in the multiplicative sense used throughout the source-to-source consistency checks. We use these recovery results as source-level sensitivity context rather than as a single hard pass/fail gate.

 \section{Results}
\label{sec:results}

\subsection{Detection Statistics}

\begin{figure}[t]
\plotone{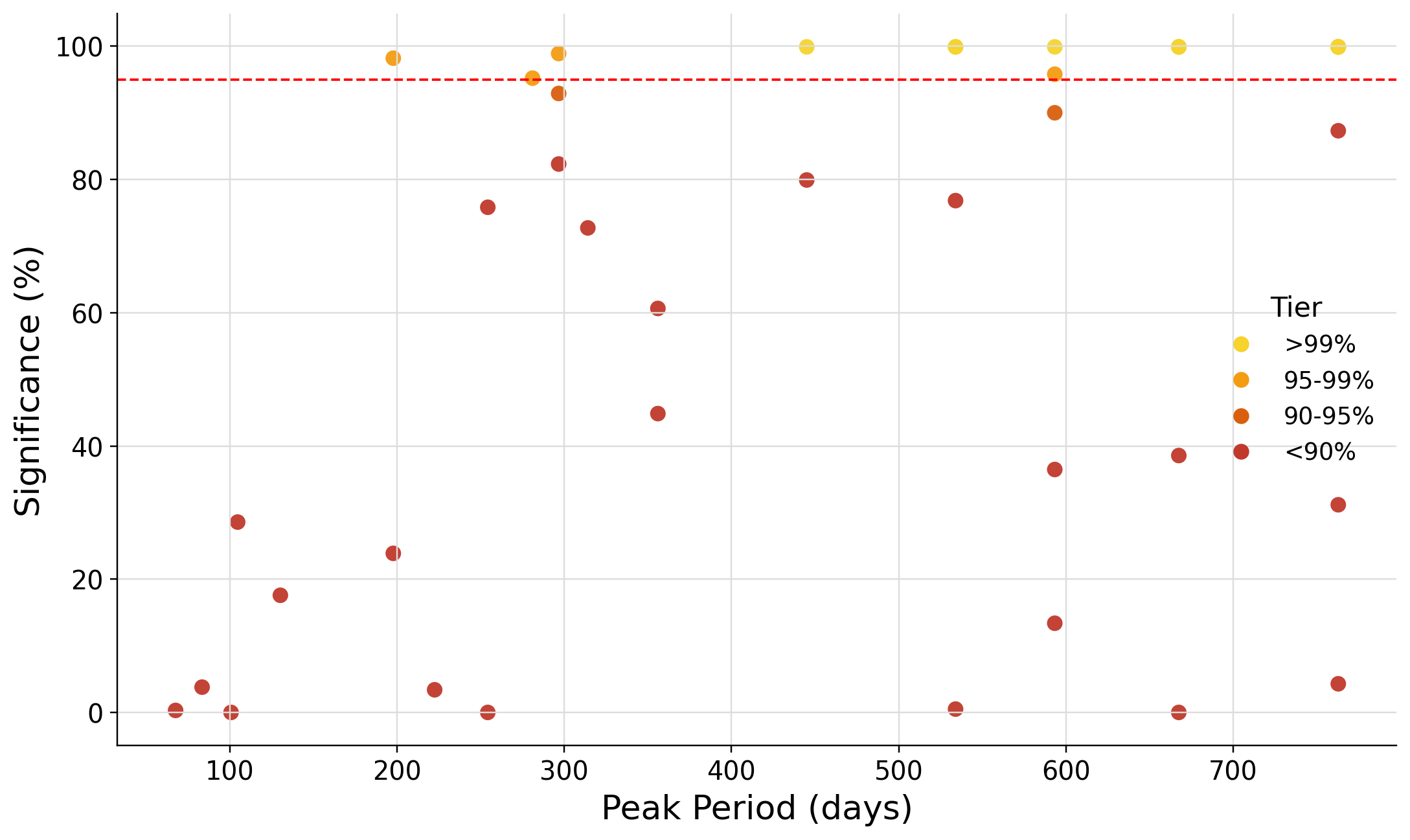}
  \caption{First-pass FFT source-level \QPO\ significance (used for candidate selection only) as a function of the detected peak period for the 41 \TeV\ blazars in the sample. Each point represents one source; the color encodes the significance tier ($>$99\% yellow, 95--99\% orange, 90--95\% dark orange, $<$90\% red). Significance values are derived from source-specific red-noise Monte Carlo tests applied to the strongest peak in the adopted 60--1000 day search range. The horizontal dashed line marks the 95\% threshold. Many sources fall below this level, and among those that exceed it, only a subset remains credible after the subsequent multiple-testing and robustness filters described in the text.}
\label{fig:sig}
\end{figure}

The multi-stage FFT pipeline identifies candidate periodicities, summarized in Table~\ref{tab:detection_summary}. Sixteen of the 41 sources exceed a raw 95\% FFT threshold, 13 remain after Benjamini--Hochberg (BH) correction, and requiring a detection fraction above 50\% and at least partial split-sample half-to-half consistency reduces the set to six strong \QPO\ candidates. These counts define a candidate shortlist; source-level significance is quantified separately with the forward-simulation test of Section~\ref{sec:corrected}, which does not rely on a periodogram slope fit. The first-pass significances in Table~\ref{tab:detection_summary} and Figure~\ref{fig:sig} come from comparing each source's strongest FFT peak with power-law red-noise simulations, and we use them only for candidate selection; all quantitative significance statements adopt the forward-simulation test of Section~\ref{sec:corrected}, with the red-noise slope treated conservatively. The detection-fraction, split-sample, Lomb--Scargle, and WWZ diagnostics used in the tiering are independent of the red-noise slope. The distribution of first-pass significance values is shown in Figure~\ref{fig:sig}.

\begin{table}[t]
\centering
\caption{Candidate-selection counts from the multi-stage FFT pipeline. Source-level significance is quantified by the forward-simulation test (Section~\ref{sec:corrected}, Table~\ref{tab:corrected}).}
\begin{tabular}{lc}
\hline
Criterion & Sources \\
\hline
Raw significance $\ge95\%$ & 16/41 \\
BH $q<0.05$ & 13/41 \\
Strong candidates & 6/41 \\
Automated multi-method candidates & 0/41 \\
Detection-only LS survivors & 2/41 \\
Ambiguous source-level cases & 2/41 \\
WWZ BH-significant & 0/41 \\
\hline
\end{tabular}
\label{tab:detection_summary}
\end{table}

The six strong candidates are J0221.1+3556, J0449.4$-$4350, J0721.9+7120, J1512.8$-$0906, J2158.8$-$3013, and J2243.9+2021. However, the formal automated multi-method tier is empty: no source simultaneously satisfies the FFT, Lomb--Scargle, and WWZ support criteria defined in the pipeline. We therefore distinguish throughout this paper between 13 statistical candidates, 6 strong candidates, and 2 ambiguous source-level follow-up cases.

The seven BH-selected sources that do not survive into the strong-candidate tier remain informative. Some are limited mainly by low detection fraction, whereas others are removed by half-to-half inconsistency between the split light curves. The final reduction is therefore not driven by a single harsh cut; different sources fail for different reasons, as expected for a mixture of comparatively stable signals, borderline cases, and red-noise features promoted by finite sampling.

The 50\% detection-fraction threshold is not uniquely privileged. When the same tier definition is evaluated at 40\%, 50\%, and 60\%, the number of strong candidates changes only from 7 to 6 to 5 sources. The contraction of the candidate set is therefore not an artifact of a single threshold tuning choice. Likewise, evaluating source-level significance under the conservative envelope over the red-noise slope described in Section~\ref{sec:methods}, rather than at a single fitted slope, does not soften the contraction of the candidate set; if anything it reinforces it, because candidates that appear significant only for a narrow range of assumed slopes do not survive the worst-case treatment.

\begin{figure*}[t]
\plotone{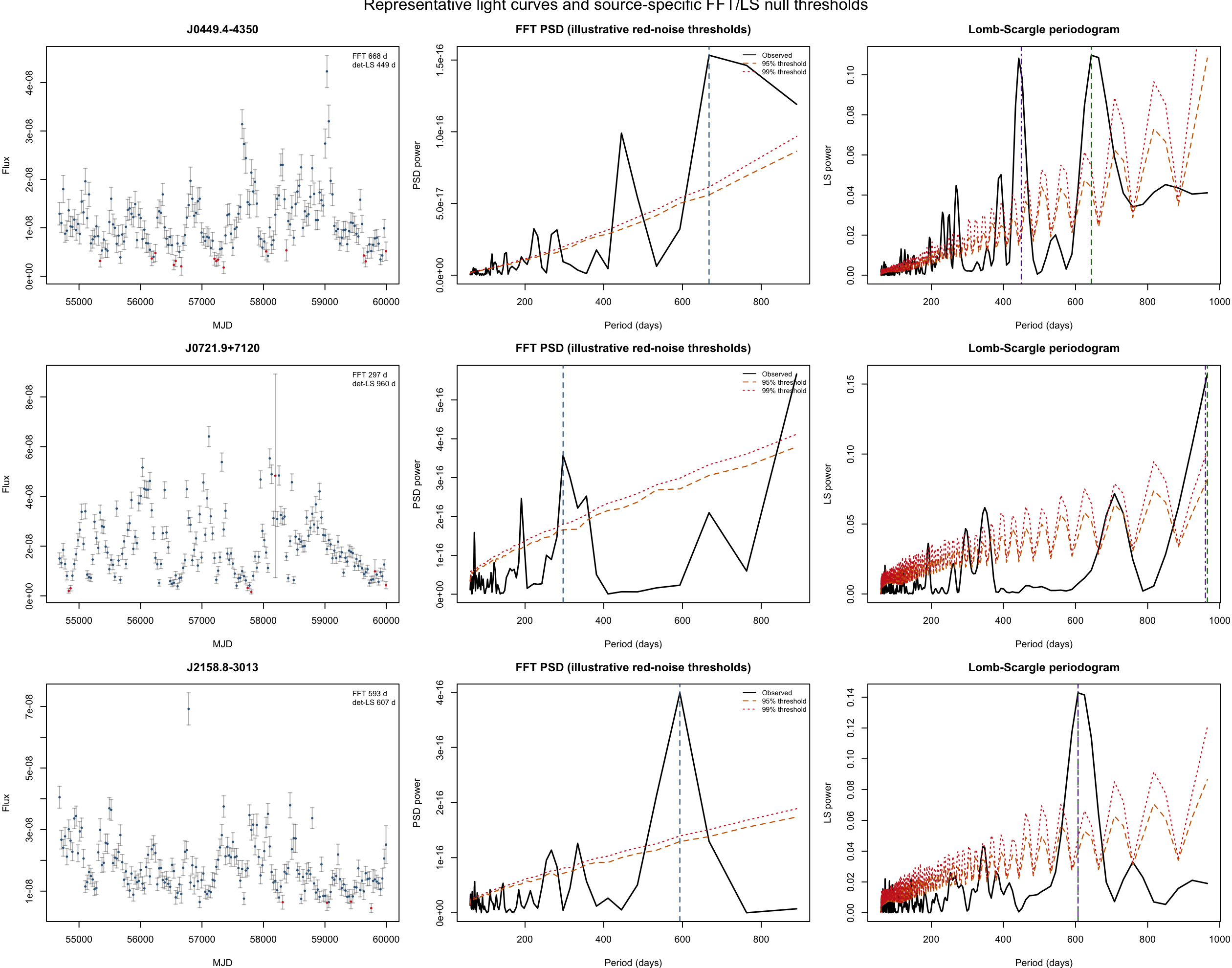}
\caption{Representative source-level diagnostics for three highlighted cases, arranged as a 3$\times$3 grid with one source per row. The left column shows the monthly \Fe\ light curves, with detected bins and lower-significance bins distinguished visually. The middle column shows the FFT-based \PSD\ over the adopted 60--1000 day search band together with illustrative source-specific 95\% and 99\% red-noise threshold curves, and the right column shows the corresponding Lomb--Scargle periodograms with the same threshold construction. These threshold curves are shown for visual orientation only; definitive source-level significance is provided by the forward-simulation test of Section~\ref{sec:corrected}. J0449.4$-$4350 (top row) illustrates a strong FFT-selected source whose auxiliary-method support weakens; J0721.9+7120 (middle row) illustrates a method-discordant case in which FFT and Lomb--Scargle favor different recurrence scales; J2158.8$-$3013 (bottom row) illustrates the most internally coherent source-level case in the sample, although it still remains provisional rather than confirmed.}
\label{fig:case_diagnostics}
\end{figure*}

Figure~\ref{fig:case_diagnostics} provides the source-level view that is missing from summary histograms alone. We highlight one stable FFT-selected strong candidate without retained detection-only support (J0449.4$-$4350), one method-discordant follow-up case (J0721.9+7120), and the most internally coherent follow-up case (J2158.8$-$3013). The figure shows explicitly that the present survey contains several qualitatively different kinds of ``interesting'' source: a prominent FFT peak is not automatically equivalent to cross-method agreement, and even the best-looking cases remain sensitive to how the null model and data censoring are handled.

 \begin{table*}[t]
\centering
\caption{Strong \QPO\ candidates from the automated workflow. The Significance and BH $q$ columns are first-pass FFT values used for candidate selection; source-level significance is given by the forward-simulation test of Section~\ref{sec:corrected}. The last column summarizes prior source-specific or source-focused \textit{Fermi}-LAT periodicity studies when we identified them in the literature. Abbreviations: Y20 = \citet{yang2020j0449}; C22 = \citet{chen2022qpo0716}; R22 = \citet{roy2022pks1510}; L23 = \citet{li2023pks1510}; Z17 = \citet{zhang2017qpo2155}; S14 = \citet{sandrinelli2014qpo2155}.}
\footnotesize
\begin{tabular}{lcccccc}
\hline
Source & Period (days) & Significance & BH $q$ & Detection rate (\%) & CV class & Prior study \\
\hline
J0221.1+3556 & 762.9 & 0.999 & 0.0034 & 60.7 & Consistent & --- \\
J0449.4$-$4350 & 667.5 & 0.999 & 0.0034 & 91.6 & Consistent & Y20 \\
J0721.9+7120 & 296.7 & 0.989 & 0.0347 & 96.1 & Partial & C22 \\
J1512.8$-$0906 & 667.5 & 0.999 & 0.0034 & 90.4 & Partial & R22; L23 \\
J2158.8$-$3013 & 593.3 & 0.999 & 0.0034 & 97.8 & Partial & Z17; S14 \\
J2243.9+2021 & 445.0 & 0.999 & 0.0034 & 51.7 & Partial & --- \\
\hline
\end{tabular}
\label{tab:strong_candidates}
\end{table*}

Table~\ref{tab:strong_candidates} uses ``CV class'' as shorthand for the split-sample cross-validation class, and ``BH $q$'' for the Benjamini--Hochberg adjusted p-value. Here the CV class summarizes agreement between the dominant periods found in the two half-light curves; it does not require either half to reproduce the full-sample period exactly.

\subsection{Source-level Significance from Forward Simulations}
\label{sec:corrected}

We quantify the source-level significance of each candidate with the forward-simulation procedure of Section~\ref{sec:methods}. For each of the 41 sources we generate $10^{4}$ \citet{emmanoulopoulos2013lightcurves} light curves on the observed detection-only sampling, evaluate a weighted (inverse-variance) sinusoidal power---a floating-mean Lomb--Scargle statistic---identically for the data and each realization, and compare the observed maximum peak power over the 60--1000 day band with the simulated maximum-peak distribution. Because the null is calibrated from complete simulated light curves passed through the same sampling window, this test assumes no independence of periodogram powers.

At the simulation-selected reference slopes, the candidate set contracts sharply. Only two of the 41 sources reach a raw source-level $p<0.05$, only J1555.7+1111 reaches BH significance ($q\simeq0.03$), and no source is significant under the more conservative Benjamini--Yekutieli correction. None of the six candidates favored by the FFT search (Table~\ref{tab:strong_candidates}) is significant: their global $p$-values range from $0.036$ for J0721.9+7120 to $0.81$ for J2243.9+2021. The strongest source, J1555.7+1111, was flagged as BH-significant by the first-pass search but did not reach the six-source strong-candidate tier, which additionally requires split-sample consistency. That a source promoted to strongest by the forward-simulation test is demoted by the first-pass filters directly illustrates how sensitive candidate rankings are to methodology.

\begin{figure*}[t]
\plotone{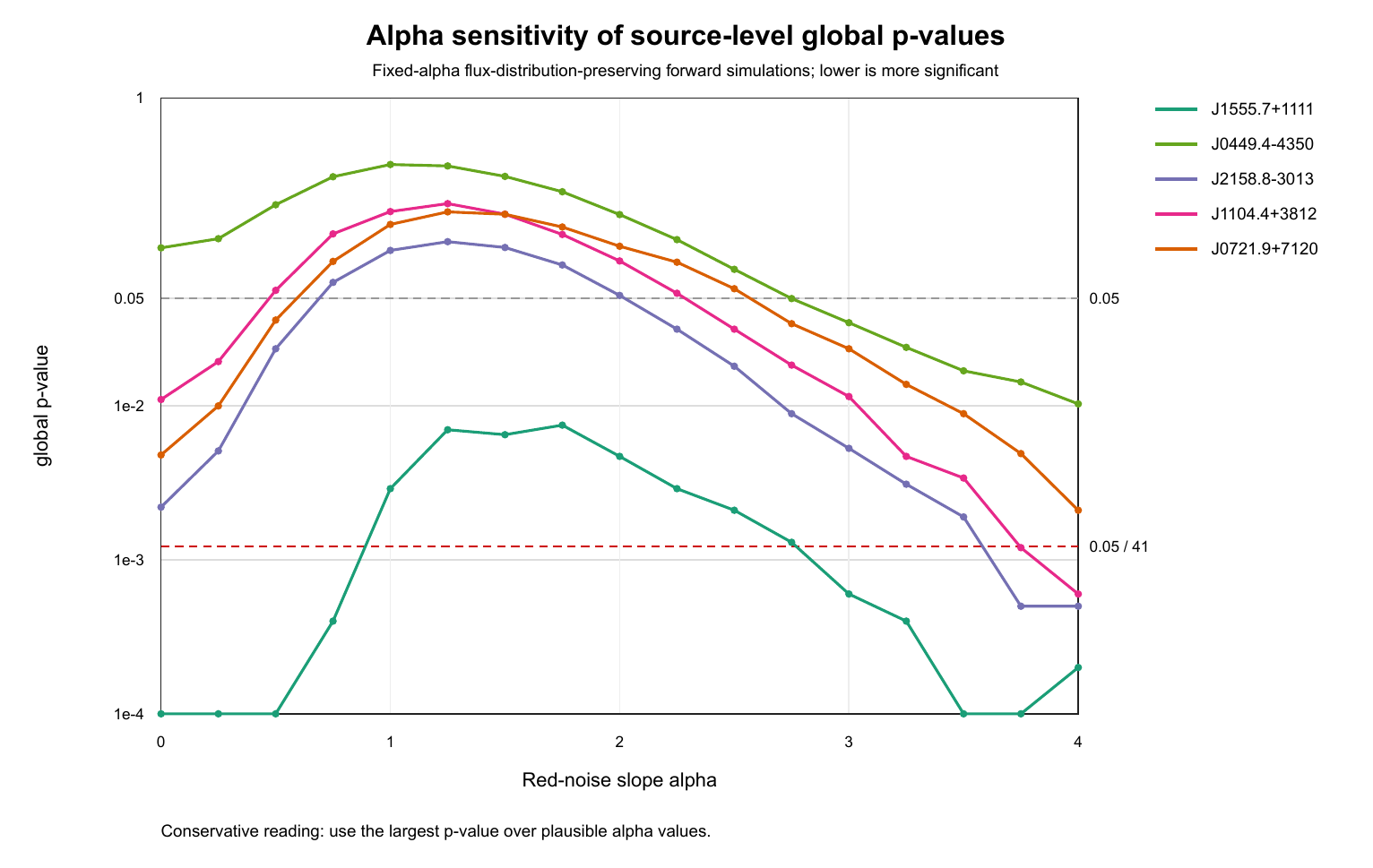}
\caption{Source-level global $p$-value as a function of the assumed red-noise slope $\alpha$ for the five lowest-$p$ sources, from fixed-$\alpha$ forward simulations that preserve the observed sampling and flux distribution ($10^{4}$ realizations per slope; lower is more significant). The dashed lines mark the nominal $0.05$ threshold and the $0.05/41$ single-trial Bonferroni level for the 41-source search. Only J1555.7+1111 remains below $10^{-2}$ across the full grid, but its conservative envelope (the largest $p$-value over the plotted range) reaches $7.5\times10^{-3}$, i.e.\ $\sim0.31$ after full-sample correction. The strong slope dependence of the significance curves motivates treating $\alpha$ as a nuisance parameter rather than basing a claim on one reference value.}
\label{fig:alpha_sensitivity}
\end{figure*}

Because the red-noise slope $\alpha$ is poorly constrained by these light curves (Section~\ref{sec:methods}), we treat it as a nuisance parameter and report the conservative envelope, i.e.\ the largest global $p$-value over $0\le\alpha\le4$. Figure~\ref{fig:alpha_sensitivity} shows the resulting $p(\alpha)$ curves for the five sources with the smallest source-level $p$-values, and Table~\ref{tab:corrected} summarizes them. The significance of every one of these sources is strongly slope-dependent. For J1555.7+1111 the global $p$-value ranges from $1\times10^{-4}$ at the most favorable slope to $7.5\times10^{-3}$ at $\alpha=1.75$. This conservative value corresponds to $41\times(7.5\times10^{-3})\approx0.31$ after the full-sample search, so J1555.7+1111 is not significant once slope uncertainty and multiple testing are both accounted for. Under the conservative envelope no source survives.

\begin{table}[t]
\centering
\caption{Forward-simulation significance for the five sources with the smallest source-level $p$-values. Here $p_{\rm ref}$ is the global $p$-value at the simulation-selected reference red-noise slope, $p_{\rm cons}$ is the conservative envelope (the largest global $p$-value over $0\le\alpha\le4$), and the last column is the 41-source Bonferroni product of $p_{\rm cons}$. No source is significant under the conservative envelope. Periods are the detection-only forward-search peaks and may differ from the FFT-interpolated periods reported elsewhere in this paper; the J0721.9+7120 peak lies at the $1000$~d upper search boundary.}
\begin{tabular}{lcccc}
\hline
Source & Period (days) & $p_{\rm ref}$ & $p_{\rm cons}$ & $41\,p_{\rm cons}$ \\
\hline
J1555.7+1111 & 792.4 & 0.0007 & 0.0075 & 0.31 \\
J0721.9+7120 & 1000.0 & 0.036 & 0.18 & $>1$ \\
J2158.8$-$3013 & 614.0 & 0.059 & 0.12 & $>1$ \\
J1104.4+3812 & 905.1 & 0.13 & 0.21 & $>1$ \\
J0449.4$-$4350 & 634.4 & 0.24 & 0.37 & $>1$ \\
\hline
\end{tabular}
\label{tab:corrected}
\end{table}

\begin{figure*}[t]
\plotone{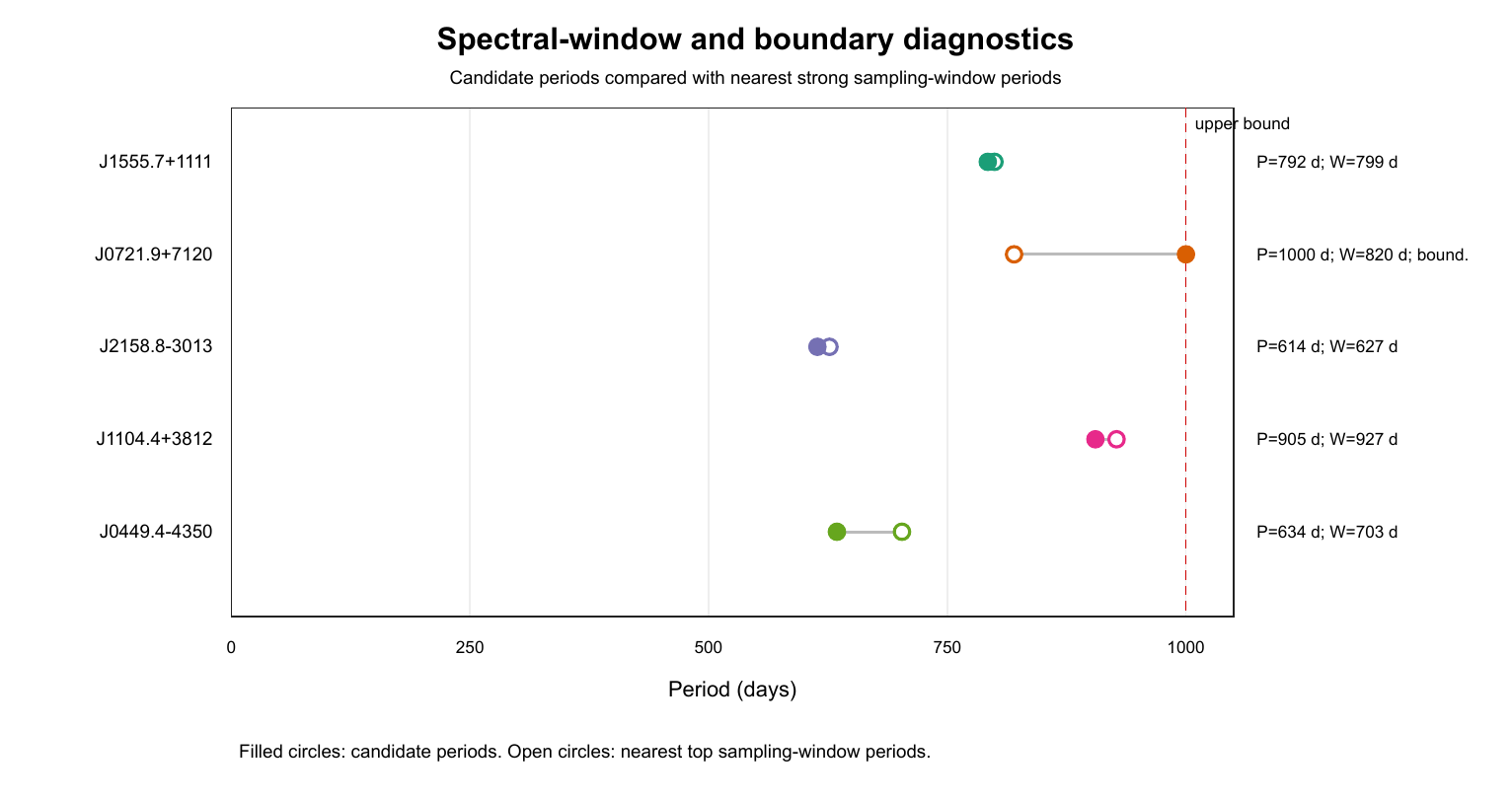}
\caption{Spectral-window and search-boundary diagnostics for the same five low-$p$ sources. Each candidate period is compared with the nearest strong feature of the detection-only spectral window. Several candidate periods lie close to weak window features, and J0721.9+7120 falls exactly on the adopted $1000$~d upper search boundary; these are reported as diagnostics, not as standalone alias tests, and reinforce the conservative classification of the candidates.}
\label{fig:window_diagnostics}
\end{figure*}

We also inspected the detection-only spectral window for these five sources (Figure~\ref{fig:window_diagnostics}). Several candidate periods lie close to weak window features, and the peak of J0721.9+7120 falls exactly on the 1000 day upper search boundary. We report these as diagnostics rather than alias rejections; they reinforce the conservative classification. Recomputing the detection mask in each of $10^4$ simulations leaves the conclusion unchanged for the three lowest-$p$ sources. At the reference and conservative-envelope slopes, respectively, the dynamic-mask global $p$-values are $3.0\times10^{-4}$ and $6.1\times10^{-3}$ for J1555.7+1111, 0.075 and 0.349 for J0721.9+7120, and 0.057 and 0.132 for J2158.8$-$3013; none is significant after the 41-source search. A pilot injection--recovery experiment for J1555.7+1111 at its conservative slope shows that the data retain $\sim50\%$ sensitivity to $0.6\sigma$ sinusoids near the candidate period and $\gtrsim80\%$ at $0.8\sigma$. The non-detection therefore reflects the strength of the red-noise continuum, not a lack of sensitivity.

 \subsection{Period Distribution and Uncertainty}

Within the strong-candidate subset, the estimated periods lie between 296.7 and 762.9 days. These values should not be interpreted as high-precision measurements. Across the full sample, the median relative bootstrap period uncertainty is 33.9\%, the mean is 60.2\%, and the largest value exceeds 300\%.

\begin{figure}[t]
\plotone{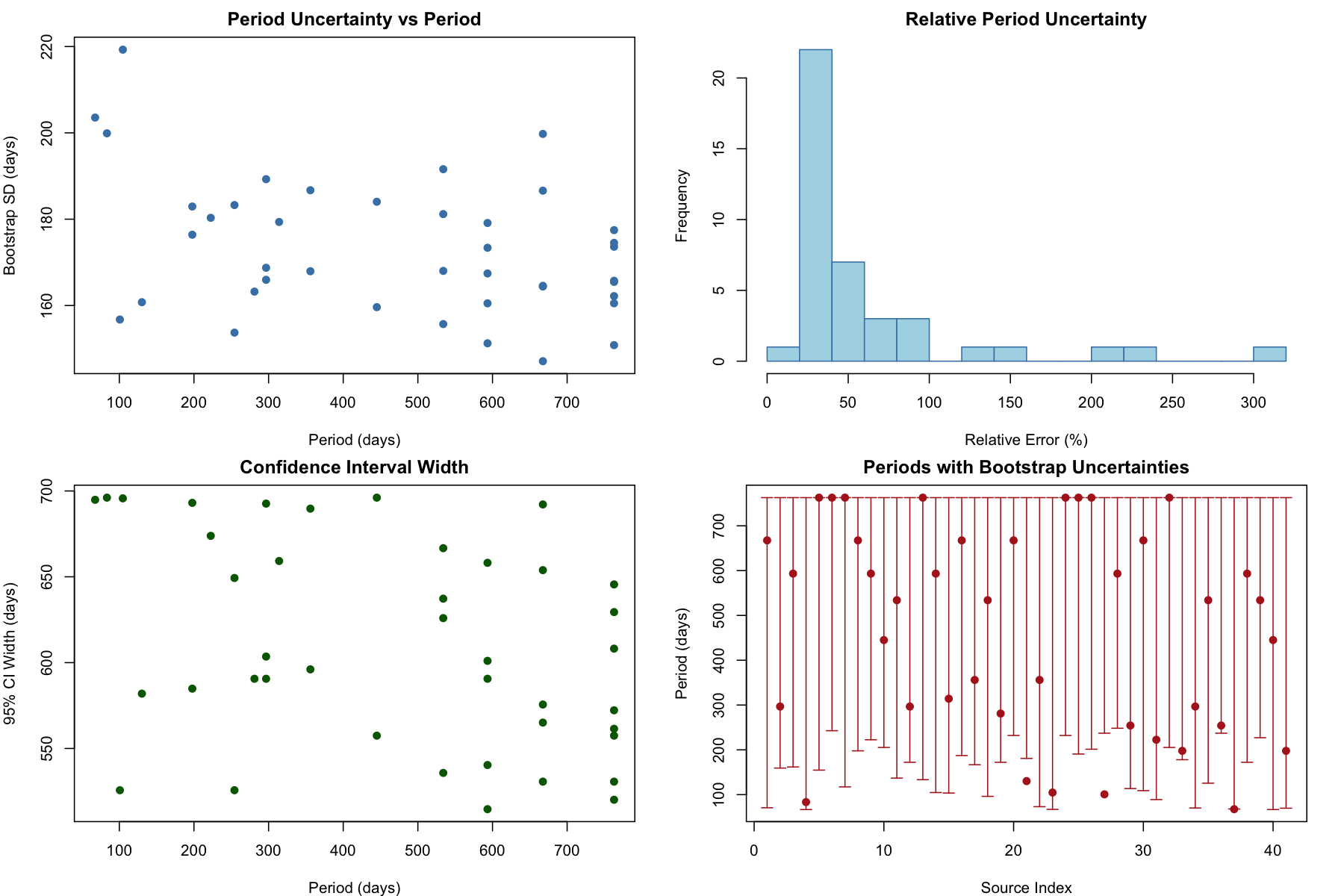}
  \caption{Bootstrap-derived period-uncertainty diagnostics for the candidate sample, shown as a four-panel figure. Top left: bootstrap standard deviation vs.\ candidate period. Top right: histogram of relative period uncertainty. Bottom left: 95\% confidence interval width vs.\ candidate period. Bottom right: individual source periods with bootstrap error bars. Together the panels show that many reported periods are only loosely localized under resampling, especially for sources with long candidate timescales relative to the finite baseline. We therefore interpret the plotted periods as characteristic recurrence scales rather than as precision measurements of stable clocks.}
\label{fig:uncertainty}
\end{figure}

Figure~\ref{fig:uncertainty} shows the bootstrap-derived period uncertainties across the candidate sample. An apparently narrow peak in a single power spectrum is therefore not sufficient to establish a stable clock. We treat the reported periods as candidate timescales unless they also pass independent robustness checks.

\subsection{Split-sample Stability}

\begin{figure}[t]
\plotone{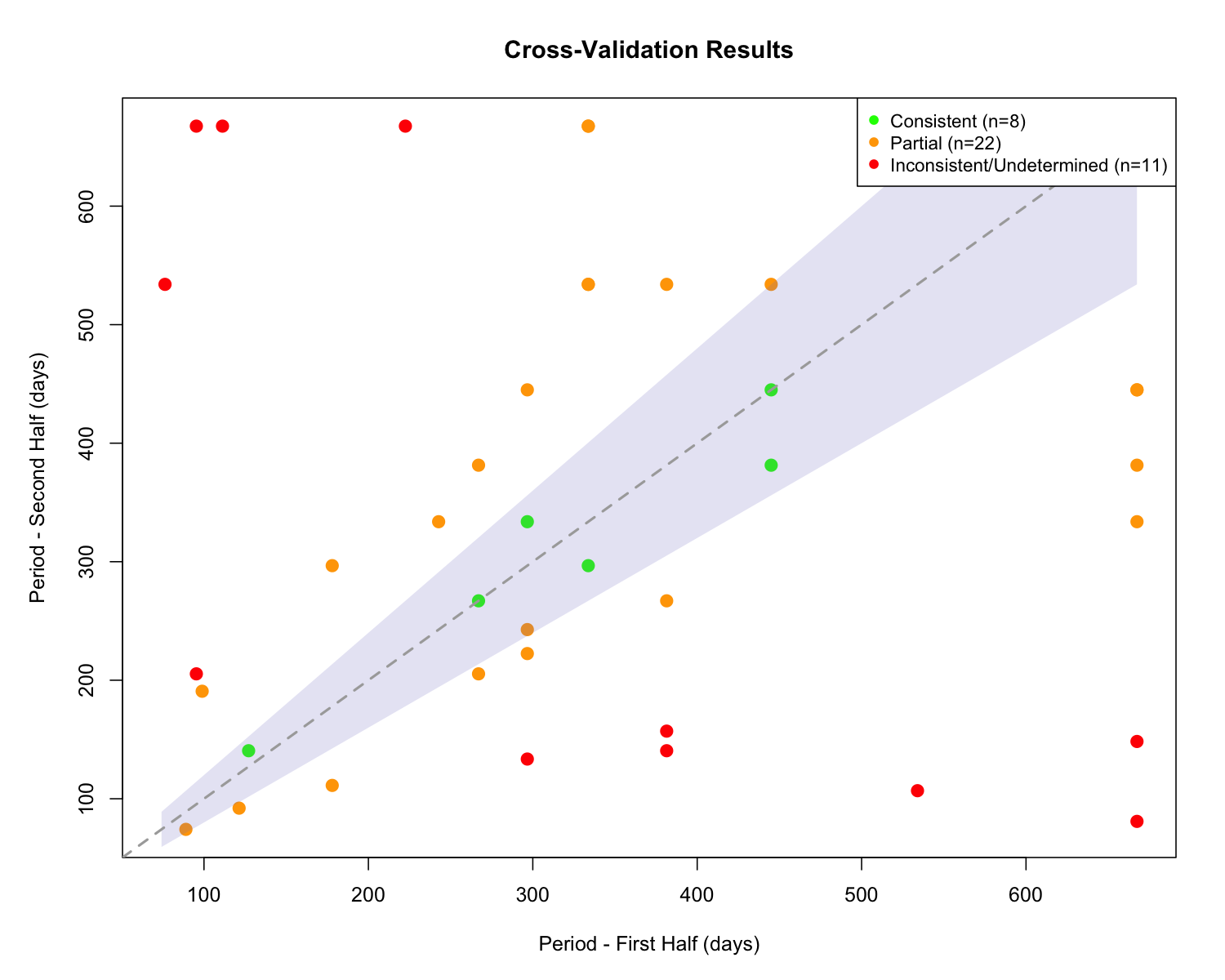}
  \caption{Split-sample cross-validation comparing the candidate periods recovered from the first and second halves of each light curve. Each point represents one source; colors denote the classification (Consistent: green, $n=8$; Partial: orange, $n=22$; Inconsistent/Undetermined: red, $n=11$). The dashed diagonal line marks the one-to-one relation and the shaded band marks the $\pm20\%$ consistency window. Sources close to the one-to-one relation are classified as Consistent, while increasingly displaced points are classified as Partial or Inconsistent/Undetermined. Only a minority of the sample falls in the tightly consistent region, reinforcing the need to treat many statistically significant peaks as provisional.}
\label{fig:cv}
\end{figure}

The cross-validation results reinforce the need for caution (Figure~\ref{fig:cv}). Across the full sample, only 8 sources are classified as Consistent, 22 as Partial, and 11 as Inconsistent/Undetermined. Even among the statistically significant subset, many sources do not reproduce the same dominant period cleanly when the light curve is split in half.

This behavior is consistent with a mixture of true but unstable quasi-periodicity, broad red-noise peaks, and limited period resolution caused by the finite baseline. The split-sample results therefore define a useful first half-to-half agreement screen, but they are not sufficient on their own.

At the same time, the split-sample test should not be overinterpreted as a fully powered stationarity diagnostic for every source. For the six strong candidates, each half-light-curve still contains roughly 3.5--9 candidate cycles, which is adequate for a coarse half-to-half screen. That screen checks whether the two halves favor similar recurrence scales; it does not require either half to peak at the exact full-sample period reported elsewhere in the paper. For weaker or longer-timescale cases, however, the same test is naturally less decisive, and an Inconsistent/Undetermined classification may reflect limited statistical power as well as genuine non-stationarity.

\subsection{Upper-limit Sensitivity and WWZ Diagnostics}

Explicitly testing the role of low-significance bins produces one of the strongest refinements in the workflow. In the strict detection-only reanalysis, upper-limit bins are removed rather than interpolated and the surviving irregularly sampled series is evaluated with Lomb--Scargle plus source-specific red-noise Monte Carlo simulations. Under this test, only two of the six strong candidates retain significant and period-consistent support: J0721.9+7120 and J2158.8$-$3013. The other four strong candidates weaken substantially once only secure detections are retained.

WWZ is even more conservative. The adopted WWZ grid spans 60--1000 days with 60 trial periods, 20 time centers, and decay constant $c=0.0125$, with the ridge-stability classes defined from the recovered stability fraction. Although several sources show Partial or Persistent WWZ stability classes, none remains globally significant after BH correction when WWZ is treated as an independent search method. In practice, WWZ is therefore most useful here as a time-frequency consistency diagnostic rather than as the primary discovery statistic. This distinction matters for interpretation: the automated multi-method tier is empty, so any source-level emphasis must be argued cautiously from converging partial evidence rather than from a formal three-method detection. The WWZ stability fraction itself should be interpreted operationally rather than physically: it measures how often the dominant ridge stays within 25\% of its own median period across the sampled time centers, not the fraction of the baseline over which a perfectly phase-stable clock is present.

\begin{figure*}[t]
\plotone{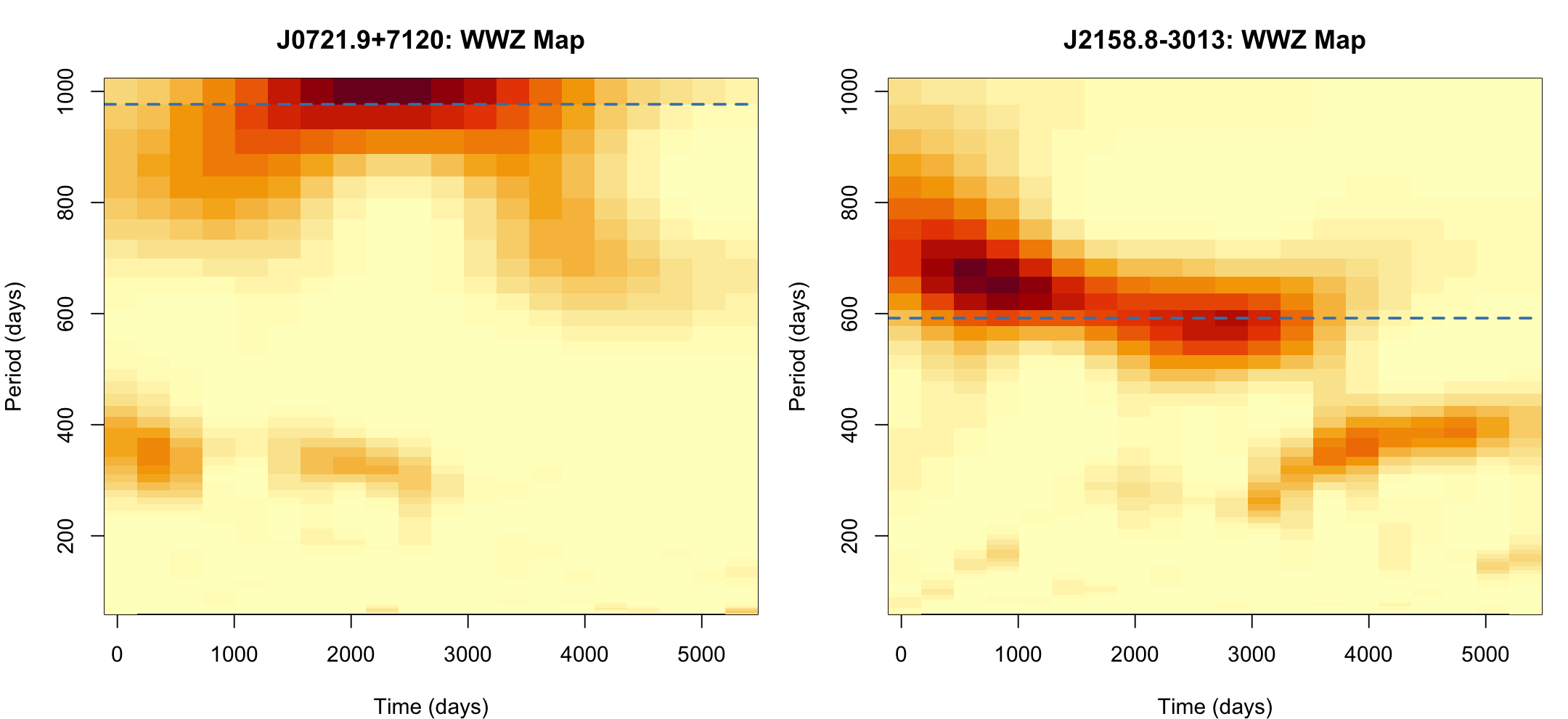}
  \caption{Representative WWZ maps for the two ambiguous source-level follow-up cases. The horizontal dashed lines mark the median period traced by the dominant WWZ ridge. J0721.9+7120 shows long-timescale structure but does not agree cleanly with the FFT-preferred period, whereas J2158.8$-$3013 exhibits a more coherent ridge near the common $\sim 600$ day timescale recovered by the different methods. These maps are used as time-frequency diagnostics rather than as stand-alone discovery figures.}
\label{fig:wwz_examples}
\end{figure*}

\subsection{Window-function and Injection--Recovery Tests}

The observation spectral window derived from the actual monthly cadence is weak across the 60--1000 day search range. Its maximum normalized power is only $3.3\times10^{-3}$, and the power sampled at the candidate periods is effectively negligible for all sources. We therefore do not use the window analysis to rank candidates positively or negatively; instead, it serves mainly to show that the cadence itself does not introduce a strong preferred scale inside the search band.

We also performed injection--recovery experiments for the two sources that survive the strict detection-only test. The simulated detection mask is recomputed from each realization rather than fixed to the observed pattern, the injected period is tied to the source-level follow-up timescale, and a successful recovery requires both period agreement and a method-specific 95\% power threshold. The injected amplitude is expressed in units of the observed light-curve standard deviation. Under this stricter definition, the mean recovery rate rises from about 0.18 (FFT), 0.10 (detection-only Lomb--Scargle), and 0.15 (WWZ) at an injected amplitude of 0.25$\sigma$ to about 0.40, 0.41, and 0.55, respectively, at 0.5$\sigma$, and reaches about 0.48, 0.87, and 0.90 at 0.75$\sigma$. The source-level behavior is heterogeneous: J0721.9+7120 remains method-discordant, so no single combined recovery metric is reported for that source, whereas J2158.8$-$3013 shows limited recovery at 0.5$\sigma$ but much stronger recovery by 0.75$\sigma$. Part of the FFT--Lomb--Scargle difference in these tests reflects the fact that FFT is evaluated on a reconstructed regular grid whereas the detection-only Lomb--Scargle branch operates directly on the surviving irregular sampling after low-significance bins are removed. For sources with many marginal bins, these two treatments weight the same underlying variability differently, so a significance mismatch between the methods is expected even when neither method is intrinsically inconsistent with the data.

\begin{figure*}[t]
\plotone{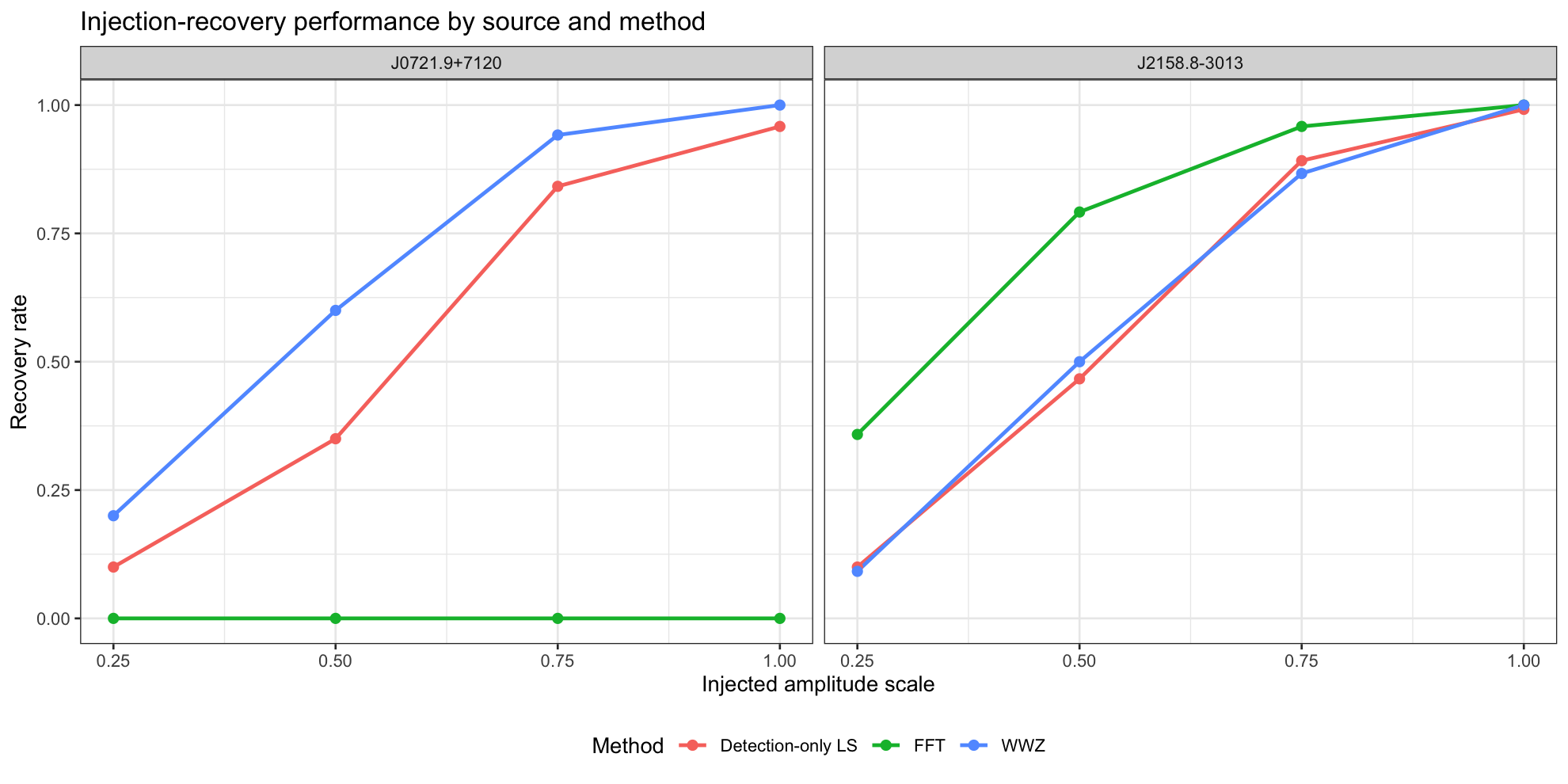}
  \caption{Injection--recovery performance for the two ambiguous source-level follow-up cases. Recovery requires both period agreement within the adopted tolerance and a method-specific 95\% significance threshold. The figure highlights the limited sensitivity of the present workflow at modest injected amplitudes, the strong method dependence of the recovery fraction, and the clear difference between the method-discordant behavior of J0721.9+7120 and the more internally coherent response of J2158.8$-$3013.}
\label{fig:injrec}
\end{figure*}

Figure~\ref{fig:injrec} summarizes the injection--recovery results for the two ambiguous source-level follow-up cases. These tests show that the effective detection efficiency is lower and more method-dependent than a simple period-matching metric would suggest. In particular, the pipeline has weak power for signals near 0.25$\sigma$ and only moderate power around 0.5$\sigma$. They also show that both cases remain uncertain even after additional source-level checks.

\begin{figure*}[t]
\plotone{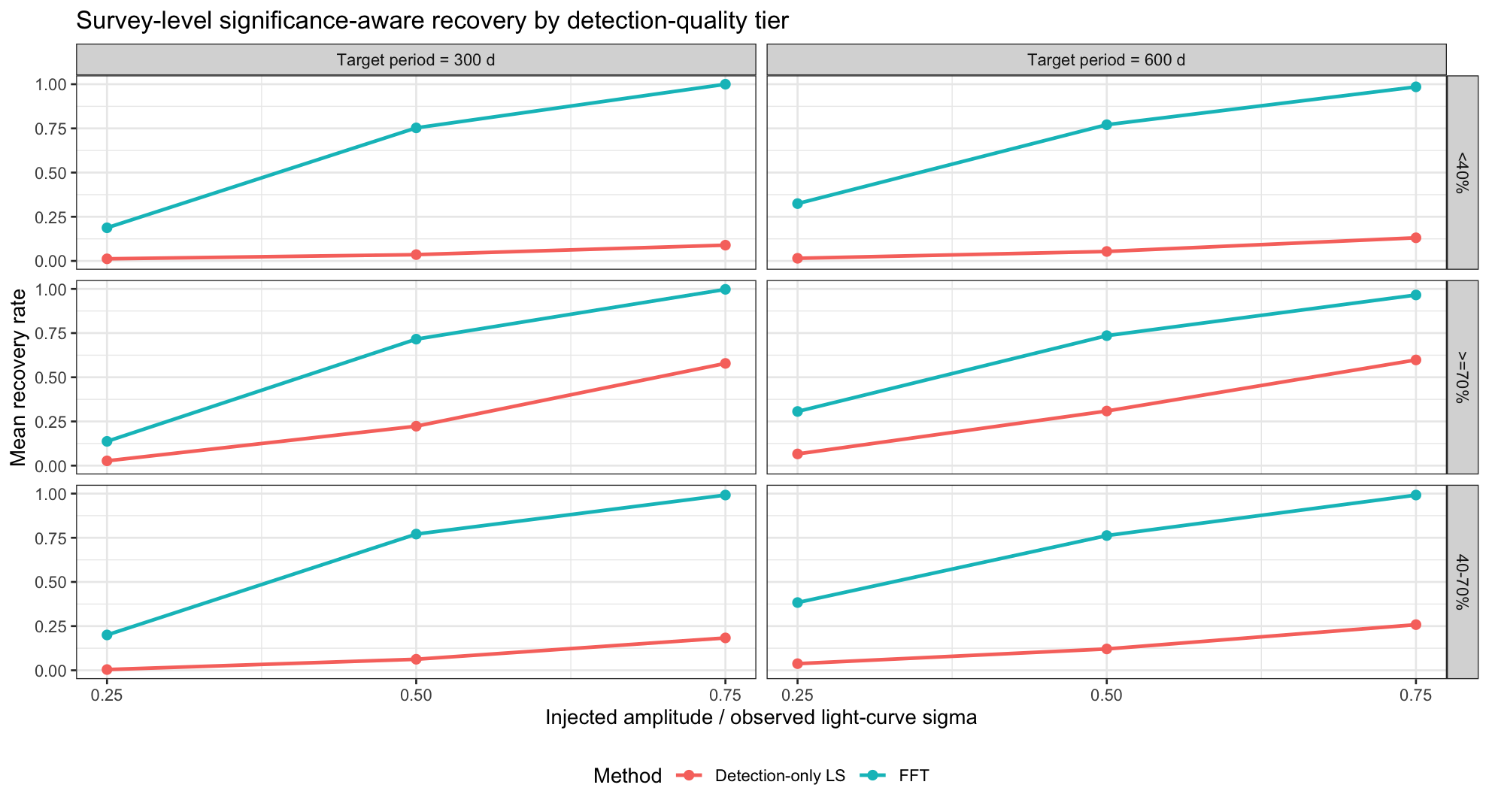}
\caption{Survey-level significance-aware recovery rates for representative injected periods of 300 and 600 days, grouped by observed detection-fraction tier. Each point averages the source-specific recovery fraction over all sources in that tier. FFT recovery becomes efficient for moderate-to-strong injected signals, whereas the stricter detection-only Lomb--Scargle criterion remains substantially less sensitive, especially in the lower-quality tiers. The figure therefore provides a sample-wide power context for interpreting the conservative null result of the automated multi-method search.}
\label{fig:survey_power}
\end{figure*}

To place these source-level tests in a broader survey context, we also carried out a sample-level significance-aware calibration across all 41 sources for representative injected periods of 300 and 600 days. Figure~\ref{fig:survey_power} summarizes the resulting recovery rates after grouping the sample by observed detection fraction. The main lesson is that the survey-wide power is heterogeneous but not pathological: FFT recovery becomes high by 0.5$\sigma$--0.75$\sigma$, whereas detection-only Lomb--Scargle remains much more demanding, especially for sources with poor or intermediate detection fraction. At 0.25$\sigma$, sample-level recovery remains low across all tiers. Even at 0.5$\sigma$, the mean detection-only Lomb--Scargle recovery is only $\sim0.05$ for sources with detection fraction below 40\%, $\sim0.09$ for the 40--70\% tier, and $\sim0.27$ when the injected period is 300--600 days in the $\ge70\%$ tier. The pipeline can therefore recover moderate-to-strong signals in favorable data, but the absence of a retained candidate in the full sample cannot be interpreted as a sharp upper limit on all possible low-amplitude quasi-periodic behavior.

\begin{table*}[t]
\centering
\caption{Ambiguous source-level follow-up cases after the automated workflow.}
\begin{tabular}{lcccccc}
\hline
Source & FFT period & LS period & Detection-only LS & WWZ median period & Window assessment & Split class \\
\hline
J0721.9+7120 & 296.7 & 965.5 & 965.5 & 976.7 & Weak window structure & Partial \\
J2158.8$-$3013 & 593.3 & 606.9 & 606.9 & 591.8 & Weak window structure & Partial \\
\hline
\end{tabular}
\label{tab:priority_candidates}
\end{table*}

\subsection{Source-level Profiles of the Detection-only Survivors}

The two detection-only survivors are listed in Table~\ref{tab:priority_candidates} and remain ambiguous after the full source-level review. J0721.9+7120 survives the strict upper-limit test and retains strong Lomb--Scargle support, but the preferred period depends strongly on the method: the FFT favors $\sim297$ days, while Lomb--Scargle and WWZ favor a much longer timescale near 960--1000 days. Together with its Partial split-sample behavior and large bootstrap uncertainty, this points to recurrent long-timescale structure rather than a well-localized periodic candidate.

The scale of this discrepancy suggests more than ordinary method scatter. Because the observing window is weak over the search range, a simple cadence-induced alias is not the preferred explanation. More plausibly, the source may contain a broad long-timescale modulation whose Fourier power is fragmented into a shorter sub-peak, or a non-sinusoidal or non-stationary structure for which FFT peak-picking emphasizes a shorter harmonic-like component while Lomb--Scargle and WWZ track the longer envelope. The present data do not allow us to discriminate cleanly among these possibilities, so we retain the source as method-discordant rather than forcing a single adopted period.

J2158.8$-$3013 is the most internally consistent case in the current sample. Its FFT, Lomb--Scargle, detection-only Lomb--Scargle, and WWZ periods are mutually consistent near 600 days, and the result survives the strict upper-limit treatment almost unchanged. Its injection--recovery behavior suggests limited sensitivity at moderate amplitude but substantially better recovery by 0.75$\sigma$. It is nonetheless not significant under the forward-simulation test ($p\simeq0.06$ at its simulation-selected reference slope, and larger under the conservative envelope; Section~\ref{sec:corrected}). We therefore retain it as the most internally consistent candidate rather than a confirmed, or even robustly significant, periodic source.

\subsection{Exploratory Source-type Context}

The feature-based clustering still supports a three-class description of variability, but it should be interpreted as a phenomenological characterization rather than a \QPO\ validation tool. The candidate set is too small to support strong claims about subtype-dependent \QPO\ behavior, and the subtype labels themselves are incomplete for part of the sample. We therefore retain source-type remarks only as exploratory context and do not use them to rank or validate candidates.

 \section{Discussion}
\label{sec:discussion}

\subsection{Interpretive Framework}

The candidate list becomes progressively smaller once data quality, temporal stability, and cross-method agreement are imposed together. Some sources are limited mainly by low-significance bins, some by inconsistent periods between halves of the light curve, and some by the absence of support from an independent method. This ranking is useful because it identifies where a candidate weakens, rather than treating all nominal detections as equivalent.

\subsection{Why the Candidate Fraction Decreases}

Several factors shape the contraction from 13 BH-selected sources to 6 strong candidates, then to zero formal automated multi-method candidates, and finally---under the forward-simulation significance test of Section~\ref{sec:corrected}---to no robustly significant source at all. First, the red-noise null model is implemented in the time domain. Second, source-level claims are tied to explicit multiple-testing correction. Third, the interpretation no longer rests on significance alone; it also depends on detection fraction, split-sample half-to-half consistency, detection-only sensitivity, cross-method consistency, and source-level recovery behavior. In this respect, the analysis follows the broader cautionary trend in red-noise-dominated variability work \citep{vaughan2016rednoise}.

This change in philosophy matters scientifically. In gamma-ray timing work, a strong Fourier peak is only the beginning of the argument. If the source spends much of the baseline near the detection threshold, if the preferred period changes substantially between halves of the light curve, or if the signal disappears when low-significance bins are masked, the appropriate interpretation is caution rather than confirmation.

The detection-fraction requirement illustrates this point. Changing the strong-candidate threshold from 50\% to 40\% or 60\% changes the count only from 6 to 7 or 5 sources, respectively. The main contraction therefore reflects the combined effect of several conservative filters rather than a single threshold tuned to force a preferred answer.

\subsection{What the Additional Tests Contribute}

The upper-limit sensitivity analysis is especially informative because it addresses a practical weakness of many survey-style gamma-ray searches. In the present sample, only two of the six strong candidates retain significant and period-consistent support when the analysis is repeated on detection-only data without reconstructing the removed bins. This distinguishes comparatively stable cases from candidates that depend more strongly on how marginal bins are treated.

The WWZ analysis adds complementary information. It identifies several sources with Partial or Persistent time-frequency behavior, but it produces no BH-significant detections when treated as an independent search method. In this sample, WWZ is more useful as a stability diagnostic than as a primary discovery statistic.

The window-function and injection--recovery tests sharpen the interpretation further. In the present monthly cadence, the observing window inside the adopted search band is weak, so it does not favor any candidate period strongly. Injection--recovery carries more weight, and in its stricter form it shows that source-level support weakens once recoveries are required to satisfy significance thresholds rather than mere period agreement. The recovery curves also show that the present workflow has limited power for modest signals: at 0.25$\sigma$ the average recovery fractions remain low for all three methods, and at 0.5$\sigma$ they are still only moderate. A separate survey-level calibration across all 41 sources reaches the same conclusion. FFT recovery becomes efficient for moderate or strong signals, but the detection-only Lomb--Scargle branch remains much more restrictive and depends strongly on the observed detection fraction.

\subsection{Implications for the Ambiguous Source-level Cases}

The remaining ambiguous source-level follow-up cases illustrate two different forms of ambiguity. J0721.9+7120 is best viewed as a source with long-timescale structure but method-dependent period assignment. The FFT prefers a shorter timescale near 300 days, whereas Lomb--Scargle and WWZ favor a recurrence scale near 1000 days. This argues against presenting the source as evidence for a sharply defined single period.

The size of the mismatch deserves explicit caution. A factor of $\sim3.3$ between the FFT and Lomb--Scargle/WWZ timescales is too large to be absorbed into the quoted bootstrap errors. Since the spectral window is weak, a simple sampling alias is not the leading explanation. More likely possibilities are that the FFT maximum is locking onto a shorter harmonic-like component of a broader modulation, or that the source contains long-timescale non-stationary structure that is summarized differently by the three methods. In either case, the prudent interpretation is that J0721.9+7120 contains interesting recurrent variability but not a uniquely defined single period in the present data.

J2158.8$-$3013 remains the strongest all-around case in the current sample. Its candidate period is consistent across FFT, Lomb--Scargle, detection-only Lomb--Scargle, and WWZ, and it survives the strict upper-limit treatment with minimal change. Even so, the split-sample classification remains Partial, the recovery experiment indicates only limited sensitivity at moderate amplitude, and it is not significant under the forward-simulation test (Section~\ref{sec:corrected}). We therefore describe it as the most internally consistent candidate rather than a confirmed or robustly significant periodic source.

The source-by-source comparison with earlier \textit{Fermi}-LAT periodicity work is also instructive. For J0449.4$-$4350, \citet{yang2020j0449} reported a marginal $\sim$450 day \QPO\ candidate, which is broadly comparable in scale to our FFT-selected 667.5 day signal but remains method-sensitive in both studies. For J0721.9+7120, \citet{chen2022qpo0716} identified a transient 31.3 day $\gamma$-ray \QPO\ in S5~0716+714 during a limited 2011--2012 interval, whereas our workflow finds a stronger recurrence scale near 297 days over the full 14.5 year monthly baseline. These results probe different timescales and observing windows, so they are better viewed as complementary evidence for non-stationary recurrent variability than as a direct contradiction. For J1512.8$-$0906, the literature is itself mixed: \citet{roy2022pks1510} reported transient $\gamma$-ray episodes near 3.6 and 92 days, whereas \citet{li2023pks1510} did not recover a comparably significant long-term \textit{Fermi}-LAT modulation, so our $\sim$668 day candidate should be interpreted cautiously. J2158.8$-$3013 has the strongest external support in the present sample. \citet{zhang2017qpo2155} reported a \textit{Fermi}-LAT recurrence near 1.74 years ($\sim$635 days), while \citet{sandrinelli2014qpo2155} found a similar $\sim$630 day $\gamma$-ray signal together with a $\sim$315 day optical timescale. Our 593.3 day candidate lies in the same broad range, which strengthens the case that this source hosts genuinely interesting year-scale variability even though the present layered tests still stop short of a firm confirmation. By contrast, for J0221.1+3556 and J2243.9+2021 we did not identify similarly focused published \textit{Fermi}-LAT \QPO\ claims, so at present they remain survey-generated candidates rather than literature-confirmed repeats.

\subsection{Implications for Physical Interpretation}

The results are most consistent with a picture in which a small number of \TeV\ blazars may host genuinely interesting recurrent gamma-ray timescales, while many others show borderline or unstable behavior. Quasi-periodic processes may occur in some \TeV\ blazars, but they are not shown to be common in the present sample. This conclusion is consistent with the broader literature, where convincing gamma-ray \QPO\ claims remain relatively rare and are usually argued on a source-by-source basis rather than as a broad population property \citep{tavani2018pg1553,sandrinelli2014qpo2155,ackermann2015qpo,zhou2018pks2247,penil2020periodicity,ren2023qpoagn}. In particular, the survey-style searches of \citet{penil2020periodicity} and \citet{ren2023qpoagn} similarly emphasize the need to compare nominal candidate peaks against a uniform statistical standard across a heterogeneous AGN sample rather than treating every local maximum as equally credible. It is also consistent with the more general caution that red-noise processes can generate apparently narrow periodogram features unless the null model and trial structure are handled explicitly \citep{vaughan2005rednoise,vaughan2016rednoise}.

At the same time, the surviving ambiguity is not astrophysically empty. Candidate recurrence timescales of several hundred days are broadly compatible with several ideas discussed in the blazar-variability literature, including geometric effects associated with helical or precessing jets \citep{camenzind1992helical,villata1999helical,liska2018jet}, longer-timescale modulation linked to the emitting flow \citep{spada2001shock}, and variability connected to the accretion system \citep{frank2002accretion}. The present data set does not allow us to discriminate among these scenarios, but the layered candidate hierarchy is valuable precisely because it identifies which sources, if any, warrant future multiwavelength or higher-cadence follow-up in that physical context.

\subsection{Practical Implications for Future Monitoring}

The same diagnostics also indicate what kind of observations would make future gamma-ray \QPO\ tests more decisive. A persuasive claim requires more than a high periodogram peak: it requires a long baseline covering several candidate cycles, a high detection fraction so that the result is not driven by upper-limit-like bins, a cadence fine enough to localize the period without moving the search close to the sampling boundary, and independent confirmation that the same timescale is recovered by methods that treat the sampling differently. In practice, the most favorable targets are bright, frequently detected blazars with candidate periods well inside the searchable range rather than near either the cadence limit or the full-baseline limit.

For observers, this argues for sustained, regular monitoring rather than short intensive campaigns alone. Higher-cadence gamma-ray binning can help only when photon statistics remain sufficient; otherwise it increases the number of low-significance bins and weakens the periodicity test. Coordinated optical, X-ray, radio, or very-high-energy monitoring is also valuable, because a recurrence timescale that appears coherently in more than one band, or persists across independent observing windows, would be less likely to reflect a single-method or single-window artifact. Future studies should therefore report detection fractions, window functions, period stability across time segments, and injection--recovery sensitivity together with any candidate period, so that non-detections and marginal cases can be interpreted as constraints on the data quality and signal amplitude rather than as simple absence of quasi-periodic behavior.

\subsection{Caveats and Limitations}

The analysis remains subject to several limitations. The monthly cadence and finite baseline limit frequency resolution and favor periods of several hundred days. The current workflow does not yet use a fully censoring-aware likelihood for low-significance bins, even though it now quantifies their impact through a detection-only sensitivity test. WWZ is used here primarily as a stability diagnostic rather than a fully optimized discovery tool, and several sources still lack secure subtype classification.

These limitations define the scope of the analysis. The present paper supports a statement about relative robustness within this 41-source sample, but it does not support a strong demographic claim about the intrinsic \QPO\ fraction of all \TeV\ blazars, nor can it uniquely distinguish among geometric modulation, jet precession, shock propagation, or accretion-related variability.

The split-sample test is a concrete example of this scope limitation. Dividing a 14.5-year light curve into two halves inevitably reduces statistical power, especially for longer candidate periods. For the six strong candidates, each half still spans roughly 3.5--9 cycles of the FFT-selected timescale, which is sufficient for a coarse half-to-half agreement screen. That screen is intentionally limited: it asks whether the two halves favor similar recurrence scales, not whether either half exactly reproduces the full-sample period. For weaker candidates or still longer recurrence scales, an Inconsistent/Undetermined split result should therefore not be interpreted as a clean proof of non-stationarity.

\subsection{Interpretive Summary}

The final picture is hierarchical. The FFT pipeline identifies 13 statistical candidates and 6 strong candidates, but no source survives the full automated multi-method tier, and the detection-only reanalysis isolates only two persistent cases (J2158.8$-$3013, the more internally consistent, and the method-discordant J0721.9+7120). The forward-simulation significance test is the decisive layer: under it no source---including these two, and including the strongest source overall, J1555.7+1111---remains significant once red-noise slope uncertainty and the full-sample search are taken into account.

 \section{Conclusions}
\label{sec:conclusions}

We have analyzed 41 \TeV\ blazars with a uniform gamma-ray timing pipeline.

\begin{enumerate}
\item A first-pass FFT search flags 16 sources at a raw 95\% level, 13 after BH correction, and six strong candidates after data-quality and split-sample filtering; none satisfies a fully automated tier requiring simultaneous FFT, Lomb--Scargle, and WWZ support.
\item Source-level significance is quantified with forward simulations that preserve the observed sampling and non-Gaussian flux distribution \citep{emmanoulopoulos2013lightcurves}, treating the red-noise slope as a nuisance parameter and never assuming independence of periodogram powers.
\item At the simulation-selected reference slopes, only two of the 41 sources reach a raw source-level $p<0.05$, only J1555.7+1111 reaches BH significance, and none survives Benjamini--Yekutieli correction. No source survives the conservative slope envelope, and none of the six FFT-selected candidates is significant.
\item J1555.7+1111 is the strongest source under this test, yet the first-pass search flagged it as BH-significant without promoting it to a strong candidate, and its global $p$-value rises to $7.5\times10^{-3}$ ($\approx0.31$ after full-sample correction) under the least favorable red-noise slope, so it cannot be claimed as a robust detection.
\item Bootstrap uncertainties, split-sample tests, and significance-aware injection--recovery experiments show limited and method-dependent sensitivity at moderate amplitudes and that many candidate periods are unstable or only loosely constrained, so candidate rankings depend strongly on methodology.
\end{enumerate}

The analysis therefore does not support the view that \QPO\ behavior is widespread across this \TeV\ blazar sample, nor does it establish any robust \QPO\ detection in the present data set: under a statistically rigorous, sampling-faithful red-noise treatment, no source remains significant once slope uncertainty and the full-sample search are accounted for. The main contribution of this paper is not a list of periodic sources but a demonstration that such a treatment substantially contracts an initially significant candidate set, providing a clearer statistical basis for evaluating the credibility of gamma-ray blazar \QPO\ candidates.

 \appendix
\section{Supplementary Candidate Tables}
\label{sec:appendix}

\subsection{Full Sample Overview}
\phantomsection
\label{sec:appendix_sample}

Table~\ref{tab:appendix_sample} lists the 41 sources analyzed in this paper together with the basic timing and detection-quality quantities used in the analysis. The purpose of this appendix table is to document the exact sample at the level of individual sources. Because uniform redshift metadata are not available for the full timing sample, the table focuses on source identity, subtype labels where available, baseline coverage, and candidate-tier bookkeeping.

\begin{deluxetable*}{lccccccc}
\tabletypesize{\scriptsize}
\tablecaption{Full 41-source sample overview.\label{tab:appendix_sample}}
\tablehead{
\colhead{Source} & \colhead{Type} & \colhead{Baseline (days)} & \colhead{$N_{\rm bins}$} & \colhead{$N_{\rm det}$} & \colhead{Detection rate (\%)} & \colhead{BH sig.} & \colhead{Strong}
}
\startdata
J0033.5$-$1921 & BCU & 5310 & 178 & 32 & 18.0 & Yes & No \\
J0035.9+5950 & HBL & 5310 & 178 & 59 & 33.1 & No & No \\
J0112.1+2245 & IBL & 5310 & 178 & 139 & 78.1 & No & No \\
J0136.5+3906 & BCU & 5310 & 178 & 107 & 60.1 & No & No \\
J0221.1+3556 & FSRQ & 5310 & 178 & 108 & 60.7 & Yes & Yes \\
J0222.6+4302 & IBL & 5310 & 178 & 163 & 91.6 & No & No \\
J0303.4$-$2407 & IBL & 5310 & 178 & 86 & 48.3 & Yes & No \\
J0449.4$-$4350 & BCU & 5310 & 178 & 163 & 91.6 & Yes & Yes \\
J0509.4+0542 & IBL & 5310 & 178 & 120 & 67.4 & No & No \\
J0521.7+2112 & HBL & 5310 & 178 & 142 & 79.8 & No & No \\
J0650.7+2503 & BCU & 5310 & 178 & 73 & 41.0 & No & No \\
J0721.9+7120 & IBL & 5310 & 178 & 171 & 96.1 & Yes & Yes \\
J0739.2+0137 & FSRQ & 5310 & 178 & 55 & 30.9 & Yes & No \\
J0809.8+5218 & HBL & 5310 & 178 & 70 & 39.3 & No & No \\
J0854.8+2006 & IBL & 5310 & 178 & 70 & 39.3 & No & No \\
J0904.9$-$5734 & IBL & 5310 & 178 & 111 & 62.4 & No & No \\
J0958.7+6534 & IBL & 5310 & 178 & 87 & 48.9 & No & No \\
J1015.0+4926 & HBL & 5310 & 178 & 168 & 94.4 & Yes & No \\
J1104.4+3812 & HBL & 5310 & 178 & 176 & 98.9 & No & No \\
J1159.5+2914 & FSRQ & 5310 & 178 &  116  &  65.2  & No & No \\
J1217.9+3007 & IBL & 5310 & 178 &  149  &  83.7  & No & No \\
J1221.3+3010 & HBL & 5310 & 178 &  57  &  32.0  & No & No \\
J1221.5+2814 & IBL & 5310 & 178 &  51  &  28.7  & No & No \\
J1224.9+2122 & FSRQ & 5310 & 178 &  70  &  39.3  & No & No \\
J1230.2+2517 & IBL & 5310 & 178 & 63 & 35.4 & Yes & No \\
J1256.1$-$0547 & FSRQ & 5310 & 178 &  159  &  89.3  & No & No \\
J1422.3+3223 & IBL & 5310 & 178 &  65  &  36.5  & No & No \\
J1427.0+2348 & IBL & 5310 & 178 &  170  &  95.5  & No & No \\
J1443.9$-$3908 & BCU & 5310 & 178 &  73  &  41.0  & No & No \\
J1512.8$-$0906 & FSRQ & 5310 & 178 & 161 & 90.4 & Yes & Yes \\
J1517.7$-$2422 & IBL & 5310 & 178 &  131  &  73.6  & No & No \\
J1555.7+1111 & HBL & 5310 & 178 & 177 & 99.4 & Yes & No \\
J1653.8+3945 & HBL & 5310 & 178 &  169  &  94.9  & No & No \\
J1725.0+1152 & BCU & 5310 & 178 &  51  &  28.7  & No & No \\
J1751.5+0938 & BCU & 5310 & 178 & 59 & 33.1 & Yes & No \\
J2000.0+6508 & HBL & 5310 & 178 &  164  &  92.1  & No & No \\
J2009.4$-$4849 & BCU & 5310 & 178 &  35  &  19.7  & No & No \\
J2158.8$-$3013 & BCU & 5310 & 178 & 174 & 97.8 & Yes & Yes \\
J2202.7+4216 & HBL & 5310 & 178 &  166  &  93.3  & No & No \\
J2243.9+2021 & BCU & 5310 & 178 & 92 & 51.7 & Yes & Yes \\
J2347.0+5141 & HBL & 5310 & 178 &  33  &  18.5  & No & No \\
\enddata
 \tablecomments{The type labels are taken from the source classifications used for this sample. ``BCU'' denotes a blazar candidate of uncertain subclass in the catalog terminology adopted for this work. $N_{\rm bins}$ is the number of monthly light-curve bins, and $N_{\rm det}$ is the number of bins that satisfy the adopted detection criterion ${\rm Flux}>3\times{\rm Error}$. ``BH sig.'' marks sources flagged by the first-pass FFT candidate-selection screen; it is not the final source-level significance assessment. ``Strong'' denotes the subset that also satisfies the adopted detection-fraction and split-sample half-to-half consistency filters in that screen. Final inference is given by Section~\ref{sec:corrected}.}
 \end{deluxetable*}

 \subsection{Statistically Significant Candidates}
\phantomsection
\label{sec:appendix_candidates}

Table~\ref{tab:appendix_candidates} lists the 13 sources flagged by BH correction in the first-pass FFT candidate-selection screen. This historical shortlist documents how the follow-up set was constructed; it must not be interpreted as 13 final detections, because source-level inference is given by the forward-simulation analysis in Section~\ref{sec:corrected}.

\begin{deluxetable*}{lcccccccc}
\tabletypesize{\scriptsize}
\tablecaption{Candidates selected by the first-pass FFT BH screen.\label{tab:appendix_candidates}}
 \tablehead{
\colhead{Source} & \colhead{Period (days)} & \colhead{Adj. confidence} & \colhead{Detection rate (\%)} & \colhead{CV class} & \colhead{Rel. error (\%)} & \colhead{Strong} & \colhead{Detection-only LS} & \colhead{Follow-up outcome}
}
 \startdata
J0033.5$-$1921 & 667.5 & $>99$ & 18.0 & Partial & 29.9 & No & No & No \\
J0221.1+3556 & 762.9 & $>99$ & 60.7 & Consistent & 22.9 & Yes & No & No \\
J0303.4$-$2407 & 762.9 & $>99$ & 48.3 & Consistent & 23.3 & No & No & No \\
J0449.4$-$4350 & 667.5 & $>99$ & 91.6 & Consistent & 24.6 & Yes & No & No \\
J0721.9+7120 & 296.7 & 95--99 & 96.1 & Partial & 55.9 & Yes & Yes & Ambiguous \\
J0739.2+0137 & 762.9 & $>99$ & 30.9 & Partial & 22.8 & No & No & No \\
J1015.0+4926 & 534.0 & $>99$ & 94.4 & Inconsistent/Undetermined  & 35.9 & No & No & No \\
J1230.2+2517 & 762.9 & $>99$ & 35.4 & Partial & 21.7 & No & No & No \\
J1512.8$-$0906 & 667.5 & $>99$ & 90.4 & Partial & 28.0 & Yes & No & No \\
J1555.7+1111 & 762.9 & $>99$ & 99.4 & Inconsistent/Undetermined  & 21.7 & No & No & No \\
J1751.5+0938 & 534.0 & $>99$ & 33.1 & Inconsistent/Undetermined  & 33.9 & No & No & No \\
J2158.8$-$3013 & 593.3 & $>99$ & 97.8 & Partial & 28.2 & Yes & Yes & Ambiguous \\
J2243.9+2021 & 445.0 & $>99$ & 51.7 & Partial & 41.4 & Yes & No & No \\
\enddata
 \tablecomments{Adjusted confidence refers only to the first-pass FFT candidate-selection screen and is superseded for inferential statements by Section~\ref{sec:corrected}. ``CV class'' denotes the split-sample cross-validation class based on agreement between the dominant periods recovered from the two half-light curves; it does not require either half to reproduce the full-sample period exactly. ``Detection-only LS'' denotes strong candidates that retain significant and period-consistent support in the strict detection-only Lomb--Scargle reanalysis. ``Follow-up outcome'' records whether a source was not carried forward or remained ambiguous after the full source-level review.}
 \end{deluxetable*}

\subsection{WWZ Configuration Summary}
\phantomsection
\label{sec:appendix_wwz}

The adopted WWZ setup is summarized in Table~\ref{tab:appendix_wwz}. We include it here to make the time-frequency configuration and stability thresholds explicit in one place, complementing the representative WWZ maps shown in Figure~\ref{fig:wwz_examples}.

\begin{deluxetable*}{lcccccc}
\tablecaption{WWZ configuration used in the manuscript.\label{tab:appendix_wwz}}
\tablehead{
\colhead{Period range} &
\colhead{Trial periods} &
\colhead{Time centers} &
\colhead{Decay constant $c$} &
\colhead{Persistent class} &
\colhead{Partial class}
}
\startdata
60--1000 days & 60 & 20 & 0.0125 & $f_{\rm stab}\ge0.7$ & $0.4\le f_{\rm stab}<0.7$ \\
\enddata
 \tablecomments{The WWZ stability fraction $f_{\rm stab}$ is the fraction of sampled time centers over which the dominant ridge remains within 25\% of the source-specific median ridge period. Representative WWZ maps for the two source-level follow-up cases are shown in Figure~\ref{fig:wwz_examples}.}
 \end{deluxetable*}

\subsection{Interpretation of the Rejected Statistical Candidates}
\phantomsection
\label{sec:appendix_rejected}

The BH-selected sources that are not classified as strong candidates fall into two broad groups. One group is limited mainly by data quality: J0033.5$-$1921, J0303.4$-$2407, J0739.2+0137, J1230.2+2517, and J1751.5+0938 all have detection fractions below 50\%, even though some of them also show Partial or Consistent half-to-half split-sample behavior. The other group is limited mainly by temporal stability: J1015.0+4926 and J1555.7+1111 have high detection fractions but Inconsistent/Undetermined split-sample periods.

Within the strong-candidate subset, the additional detection-only test removes J0221.1+3556, J0449.4$-$4350, J1512.8$-$0906, and J2243.9+2021 from the source-level follow-up list because their Lomb--Scargle support weakens once only secure detections are retained. This source-level breakdown is useful because it shows that different sources fail for physically distinct reasons rather than through a single arbitrary cut.

\section*{Data and Code Availability}
The \Fe\ monthly light curves for the 41 \TeV\ blazars are taken from \citet{wang2024}. A reproducibility package containing the input light-curve tables, the first-pass analysis scripts (\path{analysis/run_main_analysis.R} and supporting files), the authoritative forward-simulation pipeline (\path{analysis/formal_rednoise_reanalysis.py}), the dynamic-mask sensitivity check (\path{analysis/dynamic_mask_sensitivity.py}), and the key derived output products used to generate the manuscript figures and tables is available from the corresponding author upon reasonable request and will be deposited in a public Zenodo archive upon acceptance of this manuscript.

\begin{acknowledgments}
We acknowledge the Fermi-LAT Collaboration for providing the public data. This work is partially supported by the  National Key Research and Development Program of China (grant No. 2025YFA1614102), the National SKA Program of China (grant Nos. 2022SKA0120102, 2025SKA0130100), the China Space Station Telescope (CSST) project (grant No. CSST-2025-A07), National Natural Science Foundation of China (grant Nos. 12433004, 12133004, 12303019),  the Guangdong Basic and Applied Basic Research Foundation (grant Nos. 2023A1515012529, 2024A1515012218), the Innovation Research Team of Guangzhou University (grant No. 2023ZDP001). WX YANG (No.202309940006) thanks the support from Chinese Scholarship Council.
The authors used a large language model for limited language editing and organizational assistance during manuscript preparation. All scientific judgments, analysis decisions, verification, and final wording were reviewed and approved by the authors.
\end{acknowledgments}

\bibliographystyle{aasjournalv7}
\bibliography{refs}

@article{abdollahi2020fermi4fgl,
  author = {Abdollahi, S. and others},
  title = {Fermi Large Area Telescope Fourth Source Catalog ({4FGL})},
  journal = {The Astrophysical Journal Supplement Series},
  year = {2020},
  volume = {247},
  pages = {33},
  doi = {10.3847/1538-4365/ab6bcb}
}

@article{urry1995blazar,
  author = {Urry, C. M. and Padovani, P.},
  title = {Unified Schemes for Radio-Loud Active Galactic Nuclei},
  journal = {Publications of the Astronomical Society of the Pacific},
  year = {1995},
  volume = {107},
  pages = {803},
  doi = {10.1086/133630}
}

@article{padovani2017blazar,
  author = {Padovani, P. and others},
  title = {Active galactic nuclei: what's in a name?},
  journal = {Astronomy and Astrophysics Review},
  year = {2017},
  volume = {25},
  pages = {2},
  doi = {10.1007/s00159-017-0102-9}
}

@article{abdo2010psd,
  author = {Abdo, A. A. and others},
  title = {Gamma-ray Light Curves and Variability of Bright {Fermi}-detected Blazars},
  journal = {The Astrophysical Journal},
  year = {2010},
  volume = {722},
  pages = {520},
  doi = {10.1088/0004-637X/722/1/520}
}

@article{ackermann2015qpo,
  author = {Ackermann, M. and others},
  title = {Multiwavelength Evidence for Quasi-periodic Modulation in the Gamma-Ray Blazar {PG 1553+113}},
  journal = {The Astrophysical Journal},
  year = {2015},
  volume = {813},
  pages = {L41},
  doi = {10.1088/2041-8205/813/2/L41}
}

@article{timmer1995rednoise,
  author = {Timmer, J. and Koenig, M.},
  title = {On generating power law red noise},
  journal = {Astronomy and Astrophysics},
  year = {1995},
  volume = {300},
  pages = {707}
}

@article{camenzind1992helical,
  author = {Camenzind, M. and Krockenberger, M.},
  title = {The helix geometry of AGN jets},
  journal = {Astronomy and Astrophysics},
  year = {1992},
  volume = {255},
  pages = {59}
}

@article{villata1999helical,
  author = {Villata, M. and Raiteri, C. M.},
  title = {Helical jets in blazars. I. The case of Mkn 501},
  journal = {Astronomy and Astrophysics},
  year = {1999},
  volume = {347},
  pages = {30}
}

@book{frank2002accretion,
  author = {Frank, J. and King, A. and Raine, D. J.},
  title = {Accretion Power in Astrophysics},
  publisher = {Cambridge University Press},
  year = {2002},
  edition = {3rd},
  doi = {10.1017/CBO9781139164245}
}

@article{spada2001shock,
  author = {Spada, M. and others},
  title = {Internal shocks in the jets of radio-loud quasars},
  journal = {Monthly Notices of the Royal Astronomical Society},
  year = {2001},
  volume = {325},
  pages = {1559},
  doi = {10.1046/j.1365-8711.2001.04557.x}
}

@article{liska2018jet,
  author = {Liska, M. and others},
  title = {Formation of precessing jets by tilted black hole discs in 3D general relativistic {MHD} simulations},
  journal = {Monthly Notices of the Royal Astronomical Society},
  year = {2018},
  volume = {474},
  pages = {L81},
  doi = {10.1093/mnrasl/slx174}
}

@article{scargle1982,
  author = {Scargle, J. D.},
  title = {Studies in astronomical time series analysis. II. Statistical aspects of spectral analysis of unevenly spaced data},
  journal = {The Astrophysical Journal},
  year = {1982},
  volume = {263},
  pages = {835},
  doi = {10.1086/160554}
}

@article{benjamini1995fdr,
  author = {Benjamini, Yoav and Hochberg, Yosef},
  title = {Controlling the False Discovery Rate: A Practical and Powerful Approach to Multiple Testing},
  journal = {Journal of the Royal Statistical Society. Series B (Methodological)},
  year = {1995},
  volume = {57},
  pages = {289},
  doi = {10.1111/j.2517-6161.1995.tb02031.x}
}

@article{benjamini2001by,
  author = {Benjamini, Yoav and Yekutieli, Daniel},
  title = {The Control of the False Discovery Rate in Multiple Testing under Dependency},
  journal = {The Annals of Statistics},
  year = {2001},
  volume = {29},
  pages = {1165},
  doi = {10.1214/aos/1013699998}
}

@article{harris1978windows,
  author = {Harris, Fredric J.},
  title = {On the use of windows for harmonic analysis with the discrete Fourier transform},
  journal = {Proceedings of the IEEE},
  year = {1978},
  volume = {66},
  pages = {51},
  doi = {10.1109/PROC.1978.10837}
}

@article{kunsch1989bootstrap,
  author = {K{\"u}nsch, Hans R.},
  title = {The jackknife and the bootstrap for general stationary observations},
  journal = {The Annals of Statistics},
  year = {1989},
  volume = {17},
  pages = {1217},
  doi = {10.1214/aos/1176347265}
}

@inproceedings{macqueen1967kmeans,
  author = {MacQueen, J.},
  title = {Some Methods for Classification and Analysis of Multivariate Observations},
  booktitle = {Proceedings of the Fifth Berkeley Symposium on Mathematical Statistics and Probability},
  year = {1967},
  volume = {1},
  pages = {281}
}

@article{rousseeuw1987silhouettes,
  author = {Rousseeuw, Peter J.},
  title = {Silhouettes: A graphical aid to the interpretation and validation of cluster analysis},
  journal = {Journal of Computational and Applied Mathematics},
  year = {1987},
  volume = {20},
  pages = {53},
  doi = {10.1016/0377-0427(87)90125-7}
}

@article{tavani2018pg1553,
  author = {Tavani, M. and Cavaliere, A. and Munar-Adrover, P. and Argan, A.},
  title = {The Blazar {PG 1553+113} as a Binary System of Supermassive Black Holes},
  journal = {The Astrophysical Journal},
  year = {2018},
  volume = {854},
  number = {1},
  pages = {11},
  doi = {10.3847/1538-4357/aaa3f4}
}

@article{zhou2018pks2247,
  author = {Zhou, J. N. and others},
  title = {A 34.5 day quasi-periodic oscillation in $\gamma$-ray emission from the blazar {PKS 2247-131}},
  journal = {Nature Communications},
  year = {2018},
  volume = {9},
  pages = {4599},
  doi = {10.1038/s41467-018-07103-2}
}

@article{penil2020periodicity,
  author = {Pe{\~n}il, P. and others},
  title = {Systematic Search for $\gamma$-Ray Periodicity in Active Galactic Nuclei Detected by the {Fermi} Large Area Telescope},
  journal = {The Astrophysical Journal},
  year = {2020},
  volume = {896},
  number = {2},
  pages = {134},
  doi = {10.3847/1538-4357/ab910d}
}

@article{chen2022qpo0716,
  author = {Chen, Junping and Yi, Tingfeng and Gong, Yunlu and Yang, Xing and Chen, Zhihui and Chang, Xin and Mao, Lisheng},
  title = {A 31.3 day Transient Quasiperiodic Oscillation in Gamma-ray Emission from Blazar {S5} 0716+714},
  journal = {The Astrophysical Journal},
  year = {2022},
  volume = {938},
  pages = {8},
  doi = {10.3847/1538-4357/ac91c3}
}

@article{yang2020j0449,
  author = {Yang, Xing and Yi, Tingfeng and Zhang, Yan and Li, Huaizhen and Mao, Lisheng and Zhang, Haiming and Ma, Li},
  title = {The Gamma-ray and Optical Variability Analysis of the {BL Lac} Object {3FGL J0449.4-4350}},
  journal = {Publications of the Astronomical Society of the Pacific},
  year = {2020},
  volume = {132},
  number = {1010},
  pages = {044101},
  doi = {10.1088/1538-3873/ab779e}
}

@article{roy2022pks1510,
  author = {Roy, Abhradeep and Sarkar, Arkadipta and Chatterjee, Anshu and Gupta, Alok C. and Chitnis, Varsha R. and Wiita, Paul J.},
  title = {Transient quasi-periodic oscillations at $\gamma$-rays in the {TeV} blazar {PKS 1510-089}},
  journal = {Monthly Notices of the Royal Astronomical Society},
  year = {2022},
  volume = {510},
  number = {3},
  pages = {3641--3649},
  doi = {10.1093/mnras/stab3701}
}

@article{zhang2017qpo2155,
  author = {Zhang, Pengfei and Yan, Dahai and Liao, Nenghui and Wang, Jiancheng},
  title = {Revisiting Quasi-Periodic Modulation in $\gamma$-ray Blazar {PKS} 2155-304 with {Fermi} {Pass} 8 Data},
  journal = {The Astrophysical Journal},
  year = {2017},
  volume = {835},
  number = {2},
  pages = {260},
  doi = {10.3847/1538-4357/835/2/260}
}

@article{sandrinelli2014qpo2155,
  author = {Sandrinelli, Angela and Covino, Stefano and Treves, Aldo},
  title = {Quasi-periodicities of {PKS} 2155-304},
  journal = {The Astrophysical Journal Letters},
  year = {2014},
  volume = {793},
  number = {1},
  pages = {L1},
  doi = {10.1088/2041-8205/793/1/L1}
}

@article{li2023pks1510,
  author = {Li, Xiao-Pan and Cai, Yan and Yang, Hai-Yan and L\"ahtenm\"aki, Anne and Tornikoski, Merja and Tammi, Joni and Suutarinen, Sofia and Yang, Hai-Tao and Luo, Yu-Hui and Wang, Li-Sha},
  title = {Quasi-periodic behaviour in the radio and $\gamma$-ray light curves of blazar {PKS 1510-089}},
  journal = {Monthly Notices of the Royal Astronomical Society},
  year = {2023},
  volume = {519},
  number = {4},
  pages = {4893--4903},
  doi = {10.1093/mnras/stad008}
}

@article{foster1996wwz,
  author = {Foster, G.},
  title = {Wavelets for period analysis of unevenly sampled time series},
  journal = {The Astronomical Journal},
  year = {1996},
  volume = {112},
  pages = {1709},
  doi = {10.1086/118137}
}

@article{vaughan2016rednoise,
  author = {Vaughan, S. and Uttley, P. and Markowitz, A. G. and Huppenkothen, D. and Middleton, M. J. and Alston, W. N. and Scargle, J. D. and Farr, W. M.},
  title = {False periodicities in quasar time-domain surveys},
  journal = {Monthly Notices of the Royal Astronomical Society},
  year = {2016},
  volume = {461},
  number = {3},
  pages = {3145--3152},
  doi = {10.1093/mnras/stw1412}
}

@article{vaughan2005rednoise,
  author = {Vaughan, S.},
  title = {A simple test for periodic signals in red noise},
  journal = {Astronomy and Astrophysics},
  year = {2005},
  volume = {431},
  pages = {391},
  doi = {10.1051/0004-6361:20041453}
}

@article{emmanoulopoulos2013lightcurves,
  author = {Emmanoulopoulos, D. and McHardy, I. M. and Papadakis, I. E.},
  title = {Generating artificial light curves: revisited and updated},
  journal = {Monthly Notices of the Royal Astronomical Society},
  year = {2013},
  volume = {433},
  pages = {907},
  doi = {10.1093/mnras/stt764}
}

@article{ren2023qpoagn,
  author = {Ren, H. X. and Cerruti, M. and Sahakyan, N.},
  title = {Quasi-periodic oscillations in the $\gamma$-ray light curves of bright active galactic nuclei},
  journal = {Astronomy and Astrophysics},
  year = {2023},
  volume = {672},
  pages = {A86},
  doi = {10.1051/0004-6361/202244754}
}

@article{wang2024,
  author = {Wang, Gege and Xiao, Hubing and Fan, Junhui and Zhang, Xin},
  title = {{GeV Variability Properties} of {TeV Blazars Detected} by {Fermi-LAT}},
  journal = {The Astrophysical Journal Supplement Series},
  year = {2024},
  volume = {270},
  number = {2},
  pages = {22},
  doi = {10.3847/1538-4365/ad0e08}
}

@article{vaughan2010bayesian,
  author = {Vaughan, S.},
  title = {A {Bayesian} test for periodic signals in red noise},
  journal = {Monthly Notices of the Royal Astronomical Society},
  year = {2010},
  volume = {402},
  number = {1},
  pages = {307--320},
  doi = {10.1111/j.1365-2966.2009.15868.x}
}

@article{vio2010unevenly,
  author = {Vio, R. and Andreani, P. and Biggs, A.},
  title = {Unevenly-sampled signals: a general formalism for the {Lomb-Scargle} periodogram},
  journal = {Astronomy \& Astrophysics},
  year = {2010},
  volume = {519},
  pages = {A85},
  doi = {10.1051/0004-6361/201014079}
}

\end{CJK*}
\end{document}